\documentclass{aa}  

\usepackage{graphicx}
\usepackage{natbib}
\usepackage{txfonts}
\usepackage[colorlinks = true,
            linkcolor = blue,
            urlcolor  = blue,
            citecolor = blue,
            anchorcolor = blue]{hyperref}
\newcommand{\fpath}{figures/}
\begin{document}

  \title{Modelling X-shaped Radio Galaxies: Dynamical and Emission Signatures from the Back-flow model}
  \titlerunning{Modelling XRGs using Back flow model}

   \author{Gourab Giri,\inst{1}
   Bhargav Vaidya,\inst{1}
   Paola Rossi,\inst{2}
   Gianluigi Bodo,\inst{2}
   Dipanjan Mukherjee,\inst{3}
   \and Andrea Mignone\inst{4}
          }

   \institute{Department of Astronomy, Astrophysics and Space Engineering, Indian Institute of Technology Indore, Simrol, 453552, India
    \and
    INAF/Osservatorio Astrofisico di Torino, Strada Osservatorio 20, 10025 Pino Torinese, Italy
    \and
    Inter-University Centre for Astronomy and Astrophysics, Post Bag 4, Pune - 411007, India
    \and
    Dipartimento di Fisica Generale, Universita degli Studi di Torino , Via Pietro Giuria 1, 10125 Torino, Italy\\
    \email{gourab@iiti.ac.in, bvaidya@iiti.ac.in}
             }

   \authorrunning{G. Giri, B. Vaidya et al.}
 
  \abstract
   {Active galactic nuclei typically show the presence of radio jets ranging from sub-kpc to Mpc scales. Some of these radio galaxies show distortion in their jets, forming tailed or winged sources. X-shaped radio galaxies are a subclass of winged sources formation mechanism of which is still unclear.}
   {The focus of this work is to understand hydro-dynamical back-flows and their role in dynamics and non-thermal emission signatures (in presence of radiative losses and diffusive shock acceleration) during the initial phase of these galaxies.}
   {We have performed relativistic MHD simulations of an under-dense jet travelling in a tri-axial ambient using a hybrid Eulerian-Lagrangian framework to incorporate effects of micro-physical processes.}
   {We have demonstrated the dominant role played by pressure gradient in shaping XRGs in thermally dominated cases. We show that the prominence of the formed structure decreases as the jet deviates from the major axis of the ambient. The wing evolution is mainly governed by re-energized particles due to shocks that keep the structure active during the evolution time. The synthetic intensity maps of the radio galaxy show similarities with morphologies that are typically found in observed XRGs. This includes the cases with wider wings than the active lobes. The characteristic emission signatures in terms of its synchrotron spectra and implication of equipartition condition in age estimation are also discussed here. Additionally, we show that discrepancy of age can be attributed to mixing of different aged particle populations. Further, the effect of viewing angle on the difference of spectral index $(\Delta \alpha)$ of the active lobes and the wings shows a large variation and degenerate behaviour. We have demonstrated the role of diffusive shocks in the obtained variation and have concluded that the $\Delta\alpha$ spread is not a dependable characteristic in determining the formation model of XRGs.}
   {}


   \maketitle
%

\section{Introduction}

The work of \citet{Fanaroff1974} showed that there exist two types of extended radio galaxies classified based on their luminosity at 178 MHz - Fanaroff \& Riley (FR) class I and II. They can also be identified by observing the absence or the presence of the edge brightened feature at the jet termination point respectively.
The radio jets ejected along the spin axes of the central super massive black-hole (SMBH) can extend from several kpc to a few Mpc in space \citep{Bassani2016,Dabhade2020}. 
Interestingly, some of these double-lobed radio galaxies show significant distortion in their lobes and form peculiar radio structures. They can be divided into two categories -
Mirror symmetry, when the bridges or lobes bend away from the galaxy in the same direction forming tailed sources and Inversion symmetry, when they bend in opposite directions forming winged sources (X-, S- or Z-morphology).
There is a general agreement that mirror symmetry occurs when the galaxy is moving with respect to the ambient medium. In most cases, the ram pressure of the cluster medium and buoyancy forces influence the formation of these tailed structures \citep{Smolcic2007,Muller2021,Pandge2021}. 
However, the formation mechanism of X-shaped radio galaxies (or the winged sources) is still under debate with no general agreement \citep{Lal2007,Hodges-Kluck2010, Gopal-Krishna2012,Hardcastle2019,Cotton2020}. 
The first X-shaped radio galaxy, NGC 326, was discovered by \citet{Ekers1978}.
Since then, it has been known that these galaxies have a pair of active lobes similar to the classical double-lobed radio structure beside the presence of secondary lobes (wings) which are diffuse, extended and of lower luminosity than the active lobes \citep{Cheung2007,Bera2020}. A recent study using MeerKat telescope have demonstrated a classic wing-lobe structure using 1.28\,GHz radio observations demonstrating wings extending up to several hundreds of kpc \citep{Cotton2020}.

Several models have been put forward in order to describe the peculiar X-shape. One of the models is based on the merging of galaxies hosting super-massive black holes \citep{Begelman1980,Rottmann2001}. Here, the formation of diffuse lobes has been attributed to the relic jet, formed due to the BH-BH collision leading to a quick change in the jet ejection axis of the central AGN \citep{Merritt2002, Gopal-Krishna2003,Saripalli2018}. 
The process of merging of galaxies can also result in a rather slow transition of central BH spin axis. Here, the freshly deposited matter of the merged companion with different angular momentum will eventually realign the spin axis of the central BH in a gradual manner via accretion \citep{Rottmann2001}. The re-orientation time scale for such a process is of the order of $10^6-10^7$ years \citep{Natarajan1999,Dennett-Thorpe2002}. Additionally, jet re-orientation due to spin transition of the central SMBH in a post-merger phase has also been suggested by \citet{Garofalo2020}.

A different model considering the role of the ambient relates the formation of wings occurs to the evolution of the back-flowing plasma in the galaxy tri-axial density distribution   (Back-flow model). The back-flowing materials (plasma) released at the jet head while traversing back towards the centre, get deflected by the interstellar  medium (ISM) of the host galaxy leading to an X-shaped morphology \citep{Leahy1984}. In the study led by \citet{Capetti2002}, they have shown the role of over-pressured cocoon in forming XRGs and have shown a striking connection between the wing orientation and the minor axis of the ambient, confirmed by \citet{Gillone2016} on a larger sample. This is what actually has been observed for many XRGs \citep{Saripalli2009} and also has been verified in a detailed 3D simulation led by \citet{Rossi2017}. A recent evidence for the Back-flow model is provided by \citet{Cotton2020} where they have shown a double  boomerang structure supporting \citet{Leahy1984} and comparable to the results of numerical simulations by \citet{Rossi2017}. The role of an over-pressured cocoon in forming a giant radio galaxy having X-shaped morphology is also shown recently by \citet{Bruni2021}.

Some other formation mechanisms relevant to either of the model mentioned above are highlighted here. They are the dual AGN model by \citet{Lal2007}, the effect of buoyancy on the plasma bubble or lobes by \citet{Gull1973} which are getting some relevant observational support from the studies led by \citet{Lal2019,Hardcastle2019}. The role of a precessing jet in forming an X-shaped morphology can also not be ignored as shown by \citet{Horton2020} using numerical simulations of precessing jets and proxy synchrotron emission for the formed structures.
However, to date, there exists no general agreement between the proposed models to describe the formation process of these galaxies. Also, under which parametric restrictions each of these models work is still not well understood \citep{Joshi2019}. Finally, whether there exists any universal model which can explain all the properties of these galaxies or not is also an important question to ask, for these galaxies.

In our study, we will focus on the Back-flow model \citep{Capetti2002} as the possible origin of the X-morphology by using numerical simulations to study their dynamics and emission properties. In this regard, \citet{Capetti2002} and \citet{Hodges-Kluck2011} have also performed numerical simulations considering the role of the ambient and the back-flow of plasma using hydro-dynamical approach. In a more detailed simulation led by \citet{Rossi2017}, considering both the effects of magnetic field and relativistic flows, have investigated the wing formation process based on the jet and ambient parameters. However, considering the emission perspective of XRGs, these previous studies either have used proxies \citep{Hodges-Kluck2011} or have considered a simplistic treatment of it \citep{Rossi2017}. Here, we intend to model the emission aspect of these radio galaxies by properly introducing the effects of the ongoing micro-physical processes like radiative and adiabatic losses and the particle re-energization process. In this way we can follow the spectral evolution of the emitting particles and we can investigate whether re-energization processes can be present also in the wings, allowing a possible discrimination between different formation mechanisms.

We will adopt here the Eulerian-Lagrangian hybrid framework to incorporate effects due to radiation losses in presence of diffusive shock acceleration. Such hybrid models have been developed in several astrophysical codes \citep{Jones1999,Vaidya:2018,Winner2019,Huber2021A&A,Ogrodnik2021ApJS}. 
In particular, for this study we will use model of  \citet{Vaidya:2018} (see also \citet{Mukherjee2021}) in the PLUTO code \citep{Mignone2007b}. This model has recently also been used to study the influence of KH instability on kpc scale jets by \citet{borse2021}.

The paper is arranged as follows, we have discussed our numerical setup in Section \ref{Numerical setup} including both the dynamical and emission perspective of it. In Section \ref{Results: 3D runs}, we demonstrated the dynamical results obtained from different simulation runs. The associate spectral signatures obtained using the hybrid framework are discussed in detail in Section \ref{Results 3D: Synthetic spectra and the view angle input}. Further, we showcase the impact of shock acceleration on particle and spectral evolution in Section \ref{Results 3D: Particle Properties} and then summarize our work in Section \ref{Conclusions}. We have adopted here a flat cosmology with $H_0 = 69.6$ km/s/Mpc and $\Omega_m = 0.29$.

\section{Numerical setup} \label{Numerical setup}
Our numerical simulations were performed using the PLUTO code \citep{Mignone2007b} which solves the set of relativistic magneto-hydrodynamic (RMHD) equations defined as
\begin{equation}
\begin{split} \label{eq:1}
    \partial_{\kappa} (\rho u^{\kappa}) = 0\\
    \partial_{\kappa} (w u^{\kappa} u^{\delta} - b^{\kappa} b^{\delta} +p g^{\kappa \delta}) = 0\\
    \partial_{\kappa} (u^{\kappa} b^{\delta} - u^{\delta}b^{\kappa}) = 0
\end{split}
\end{equation}
where $\kappa$, $\delta$ $= (0,1,2,3)$ \citep{Mignone2006,Mignone2009}. Here $\rho$ is the rest mass density, $u^{\kappa}$ is the four velocity ($\Gamma(1, \textbf{v})$: $\Gamma$ represents the jet Lorentz factor) and $b^{\kappa}$ is the covariant magnetic field defined through the lab field \textbf{B} as ($b^0 , \textbf{B}/\Gamma + b^0 \textbf{v}$). Total enthalpy and total pressure of the system are represented by $w$ and $p$ respectively. The metric $g^{\kappa \delta}$ is considered to be flat having the form diag(-1, 1, 1, 1).

\subsection{Dynamical model}
The ambient condition of our numerical model is a representative tri-axial ellipsoidal galaxy. We have initialized such a galaxy using the King's density profile \citep{Cavaliere1976} which is defined (in Cartesian system) as 
\begin{equation} \label{eq:2}
    \rho = \frac{\rho_{0}}{\left(1+x'^{2}/a^{2}+y'^{2}/b^{2}+z^{2}/c^{2}\right)^{\frac{3}{4}}}
\end{equation}
where $\rho_{0}$ is the central density of the galaxy and set to be 1 amu/cc. The values of $a$, $b$, $c$ (effective core radius) are different so that the density distribution acquires a tri-axial shape. Here
\begin{equation}
    x' = x{\rm{cos}}\psi - y{\rm{sin}}\psi,\ \ y' = x{\rm{sin}}\psi + y{\rm{cos}}\psi
\end{equation}
and $\psi$ is the angle between the jet ejection axis and the major axis of the galaxy \citep{Rossi2017}.
The pressure distribution $P(x', y', z)$ follows the density distribution of the galaxy by assuming an isothermal atmosphere and is used to obtain acceleration due to gravity (\textbf{g}) through the hydro-static equilibrium equation, $\nabla P = \rho \textbf{g}$. 
Further, we have introduced an initial jet nozzle at the centre of this distribution which will continuously inject relativistic jet having Lorentz factor $\Gamma = 5$. The injected jet materials are under-dense with a factor ($\eta$) of $10^{-6}$ relative to the central density of the galaxy.
The galaxy is initially iso-thermal and the temperature of the medium is fixed by setting the jet Mach number to 400. 
Taub-Matthews equation of state has been used for our runs \citep{Taub1948, Mignone2005}. We have performed these simulations with second order accuracy in space using linear reconstruction and HLLC Riemann solver. 
We have also incorporated an initial toroidal B-field in the jet injection region which is defined as 
\begin{equation}
B_{x} = -B_{t}r{\rm sin(\vartheta)},\ \ 
B_{y} = B_{t}r{\rm cos(\vartheta)}
\end{equation}
where $r$ and $\vartheta$ are the polar coordinates in the perpendicular plane to the jet ejection axis and $B_x$, $B_y$ are the components of the field defined in that plane \citep{Rossi2017}. The value of $B_{t}$ is constant and is determined through the jet magnetization parameter $\sigma$ (set to 0.01) which is the ratio of Poynting flux to the matter energy flux ($\frac{B_t^2}{\Gamma^2 \rho h};\ h$ is the specific enthalpy). To ensure the solenoidal condition of magnetic field, we have adopted the method of divergence cleaning as prescribed by \cite{Dedner2002}. 
The simulations were conducted in dimensionless units, where the units for their conversion back into physical forms are chosen by keeping consistency with the observational data.
The unit length is defined as $L_0^{\rm 3D} =$ 4 kpc which is a fictitious core radius of the galaxy. Based on this, all the lengths are defined in our simulations.
Unit density ($\rho_0$) and velocity ($V_0$) are chosen as 1 amu/cc and light speed ($c$) respectively, using which we can find out the units for other variables as well.

\begin{figure}
\centering
\includegraphics[width=\columnwidth]{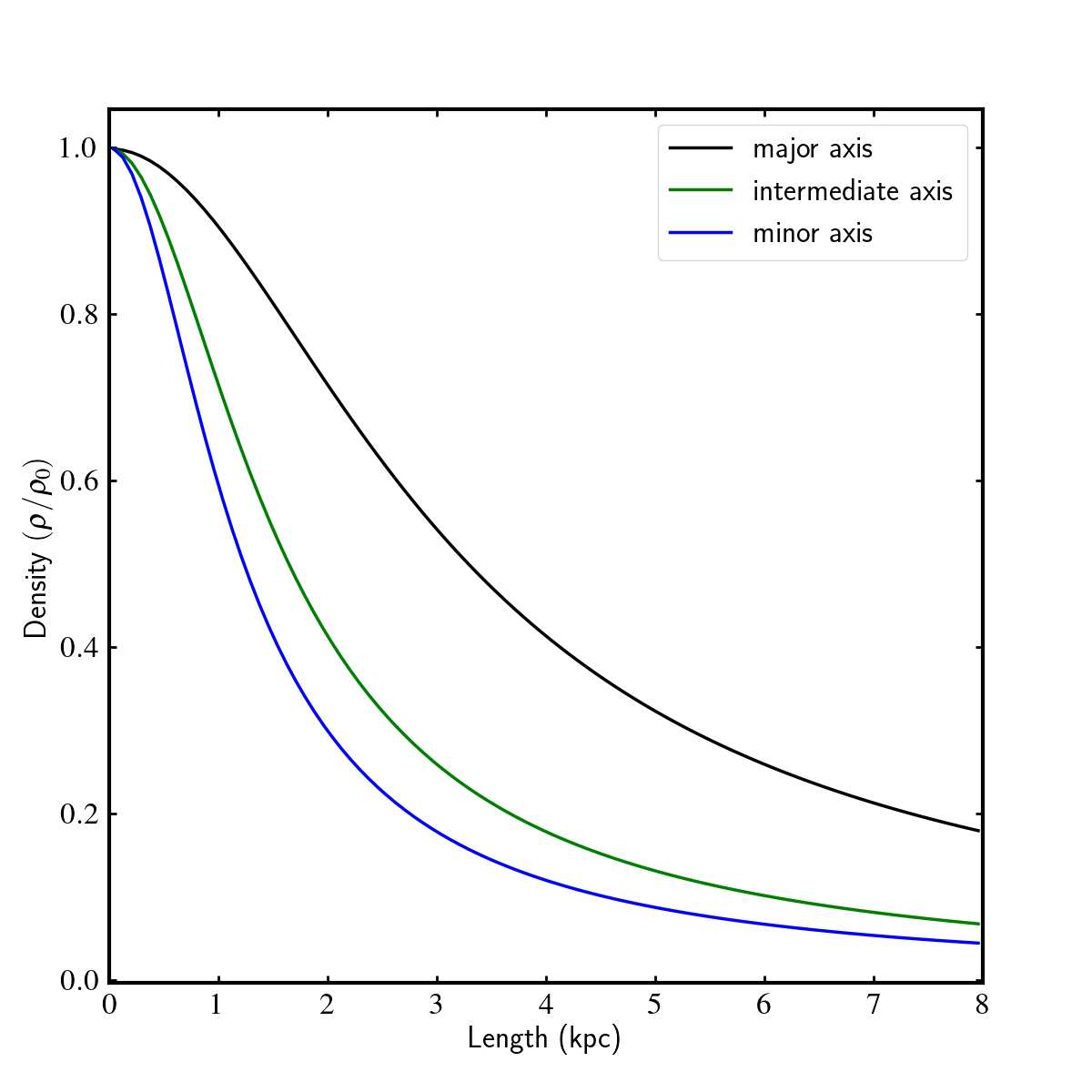}
\caption{Distribution of density along the major, intermediate and minor axis of the tri-axial ellipsoidal galaxy is shown here (in the initial phase), where the value of $\rho_0$ is 1 amu/cc.}
\label{Fig:Density_variation}
\end{figure}

\indent For our runs, we have particularly followed the initial configuration described in \cite{Rossi2017}. The initial density profile is obtained using Eq. (\ref{eq:2}) by adopting $a = 1/3$, $b = 2/3$ and $c=1/4$ with respect to the physical scale $L_0^{\rm 3D}$.
So, $y'$-axis is the major axis of the galaxy (where $y$ is the jet ejection axis) and $x'$, $z$ axes are the intermediate and minor axis respectively. The distribution of density along the three axes of the ambient medium is represented in Fig. \ref{Fig:Density_variation}.
The simulation domain has a physical size of $4L_0^{\rm 3D}$ $\times$ $8L_0^{\rm 3D}$ $\times$ $4L_0^{\rm 3D}$ which is discretised using 192 $\times$ 384 $\times$ 192 grid points for our reference runs (see Table~\ref{Tab:Parametric_space}).
A bi-directional jet is injected into the domain from the origin through a cylindrical injection region having a radius of $0.05$ and a height of $0.063$ (with respect to $L_0^{3D}$). This region  is further divided (along $\hat{y}$) into three sections - neutral region, jet region (towards +ve $y$-axis) and the counter-jet region (towards -ve $y$-axis).  
In the neutral region ($|y| <  0.02\ L_0^{3D}$), the values of B-field and velocity are zero but in the jet and the counter-jet region, the direction of the velocity and B-field are reversed. The parameters defined inside the jet injection region are set to be fixed with time. For these runs, the outer boundaries on all sides are set by extrapolating the ambient values.  

The injected jet power is calculated using the formula below \citep{Matthews2019}
\begin{equation}
   Q_j = \pi r_j^2 {\rm v}_j \bigg[ \Gamma (\Gamma - 1)\rho_j c^2 + \frac{\rotatebox[origin=tr]{90}{$\prec$}}{\rotatebox[origin=tr]{90}{$\prec$} - 1}\Gamma^2 P_j\bigg] 
\end{equation}
where $r_j$, ${\rm v}_j$, $\rho_j$, \rotatebox[origin=tr]{90}{$\prec$} and $P_j$ are representing the jet radius, bulk velocity of the jet, jet density, adiabatic index and jet pressure respectively. The values of $Q_j$, obtained for all of our runs are the same and is $1.51 \times 10^{45}$ erg/sec. 

The different simulations carried out in the present study are summarized in Table~\ref{Tab:Parametric_space}.
We have performed two principle runs in 3D using the resolution prescribed above, which we refer to as our reference runs. For the case $3Dmaj\_ref$, the ejected jet propagates along the major axis of the galaxy and for the case $3Dang\_ref$, the jet propagates at an angle $30^{\circ}$ to the major axis of the galaxy \citep[similar to cases E and F of][]{Rossi2017}. We have also carried out a high resolution run ($3Dmaj\_high$) with 384 $\times$ 768 $\times$ 384 grid cells to investigate the resolution effect in our results. 

\begin{table}
\caption{This table highlights those characteristics of our simulations which we have varied. In the 1st column, we have labeled these runs for further references. The 2nd column is representing the angle between the ejected jet and the major axis of the galaxy. The corresponding domain size of these runs (in units of 4 kpc ($L_0^{\rm 3D}$)) with the number of grid cells that enclose the domain are highlighted in column 3 and 4 respectively.}
\begin{center}
\begin{tabular}{ r|c|c|c } 
 \hline
 Sim. label&Angle-$\psi$ & Domain ($\times \ L_0^{\rm 3D}$) & Grid\\
 \hline
 $3Dmaj\_ref$&$0^{\circ}$& $4\times8\times4$ & $192\times384\times192$\\ 
 $3Dmaj\_high$&$0^{\circ}$& $4\times8\times4$ & $384\times768\times384$\\ 
 $3Dang\_ref$&$30^{\circ}$& $4\times8\times4$ & $192\times384\times192$\\ 
 \hline
\end{tabular}

\label{Tab:Parametric_space}
\end{center}
\end{table}

\subsection{Emission model} \label{Emission model}
\begin{figure}
\centering
\includegraphics[width=\columnwidth]{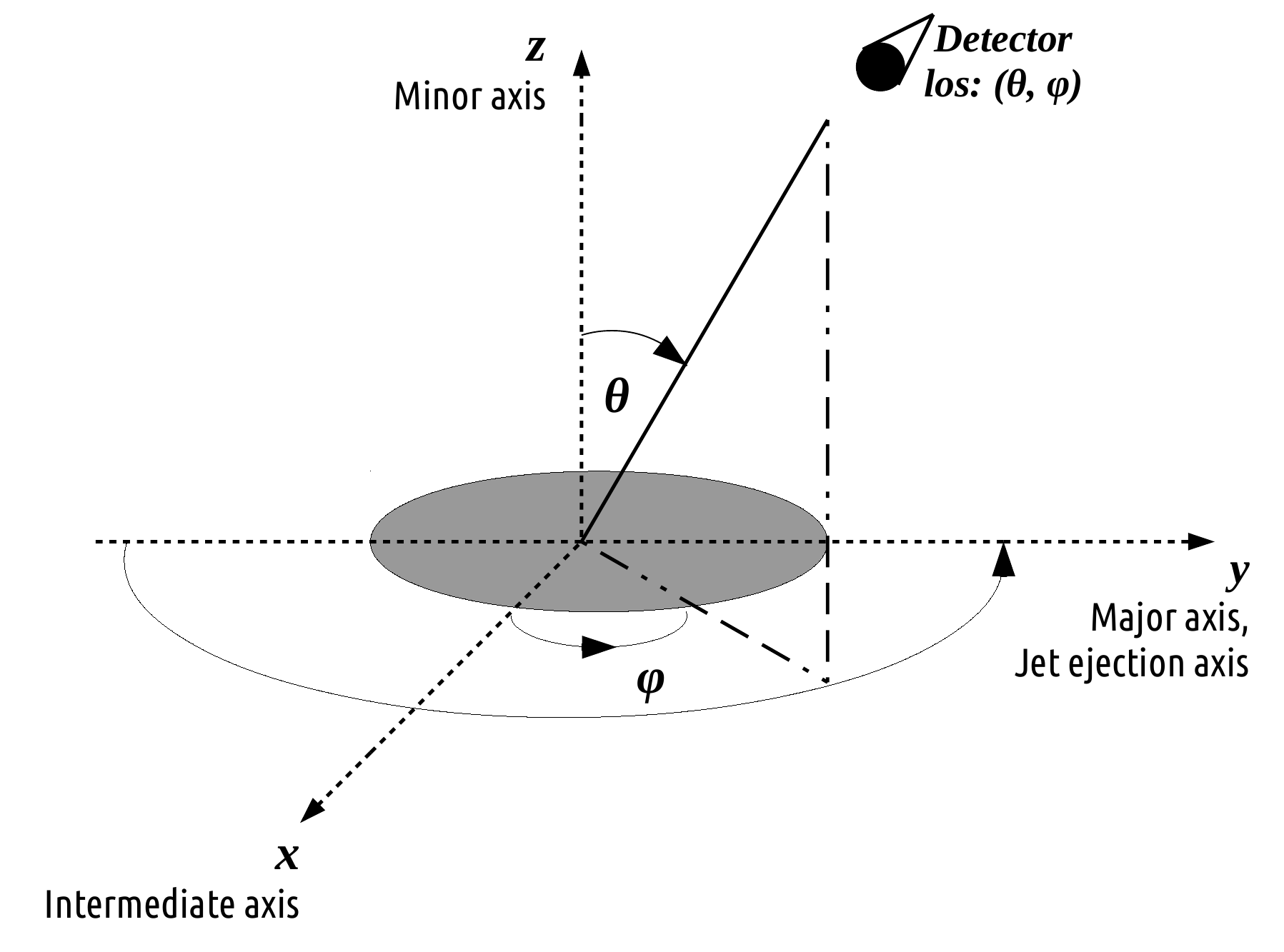}
\caption{A 3D representative image showing the line of sight visualization angle ($\theta,\ \phi$) used in our study to get the synchrotron emissivity (and hence the intensity) along that direction. The major, intermediate and minor axes of the central ellipsoidal galaxy are labelled here just to represent the relative configuration of the tri-axial ambient, where they shift slightly based on our different runs.}
\label{Fig:cartoon}
\end{figure}

We make use of the hybrid framework of PLUTO code in order to inject the Lagrangian macro-particles into the domain to model the emission properties of these galaxies \citep{Vaidya:2018,Mukherjee2021}. These non-thermal particles are continuously injected through the jet nozzle in order to fill the computational domain. With the chosen injection rate, we obtained nearly $10^6$ particles at the end of the simulation which can adequately model the emission features. Each macro-particle is an ensemble of non-thermal electrons which are set to follow a power law pattern (initially) defined as $N(\gamma) = N_0 \gamma^{-p}$ with index $p=6$. 
The value of $N_0$ can be obtained by $\int_{\gamma_{\rm min}}^{\gamma_{\rm max}} N(\gamma) d\gamma = n_{\rm micro}$ where $n_{\rm micro}$ is the non-thermal electron number density. Its value is set by assuming that the energy density of the injected electrons ($U_e$) is a fraction of the magnetic energy density $\left( \frac{B_{\rm dyn}^2}{8\pi}\right)$ at the injection region, i.e. (in cgs units)  
\begin{equation} \label{eq}
U_e = m_e c^2 \ \int_{\gamma_{\rm min}}^{\gamma_{\rm max}} \gamma N(\gamma) d\gamma \left(  = \frac{B_{\rm eq}^2}{8\pi} \right)= \epsilon \frac{B_{\rm dyn}^2}{8\pi}
\end{equation}
The term in the bracket in Eq. (\ref{eq}) represents the equipartition magnetic energy density \citep{Hardcastle2002} where $B_{\rm eq}$ is the equipartition magnetic field strength. In reality, $B_{\rm eq}$ is rather a proxy for magnetic field defined such that magnetic energy associated is equivalent to the radiating electrons energy and therefore cannot be regarded as physical magnetic field.
Here, $\epsilon$ represents the fraction and is set to the value of $2.7 \times 10^{-4}$ under the limiting Lorentz factors $10^2$ ($\gamma_{\rm min}$) and $10^{10}$ ($\gamma_{\rm max}$) of the electrons. The choice of $\epsilon$ is rather arbitrary and results in a sub-equipartition strength of emission with $B_{\rm eq} =  0.01B_{\rm dyn}$. We therefore obtain the value of $n_{\rm micro}$ as $10^{-3} \rho_j$  where $\rho_j$ is the jet density. 
The initial spectral distribution of the electrons will subsequently get updated depending on the micro-physical processes it undergoes like radiative losses, adiabatic cooling or diffusive shock acceleration from multiple shocks. For this reason, the particles are insensitive to the initial conditions when they travel some distance away from the injection region. The radiative losses considered here are the synchrotron and the inverse Compton (IC-CMB) emission, for which we set a representative redshift ($z$) of the galaxy as 0.05.
For the present study, we have focused on the synchrotron emission for seven different frequencies, six of which are in radio bands (LOFAR 144 MHz, GMRT 240 MHz, GMRT 610 MHz, VLA 1.4 GHz, VLA 5 GHz and VLBA 43 GHz) and one is in optical B-band (HST $6.5\times10^4$ GHz). In this regard, we have obtained specific intensity maps of the radio galaxy from synchrotron emissivities (at these frequencies) assuming a line of sight angle ($\theta,\ \phi$) as shown in Fig. \ref{Fig:cartoon}. 

\section{Dynamical Evolution : Lobe and Wing Formation} \label{Results: 3D runs}
In this section, we describe the dynamical evolution of various simulations leading to the formation of X-shaped morphology. 
\subsection{Jet along major axis}
\begin{figure*}
\centering
\includegraphics[width=8.6cm, height = 5.22 cm]{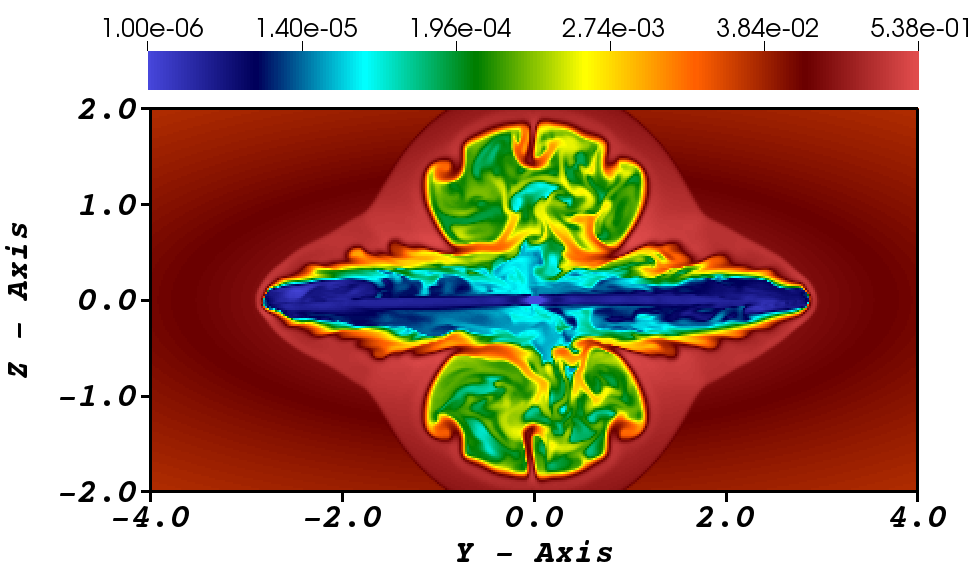}
\includegraphics[width=8.6cm, height = 5.5 cm]{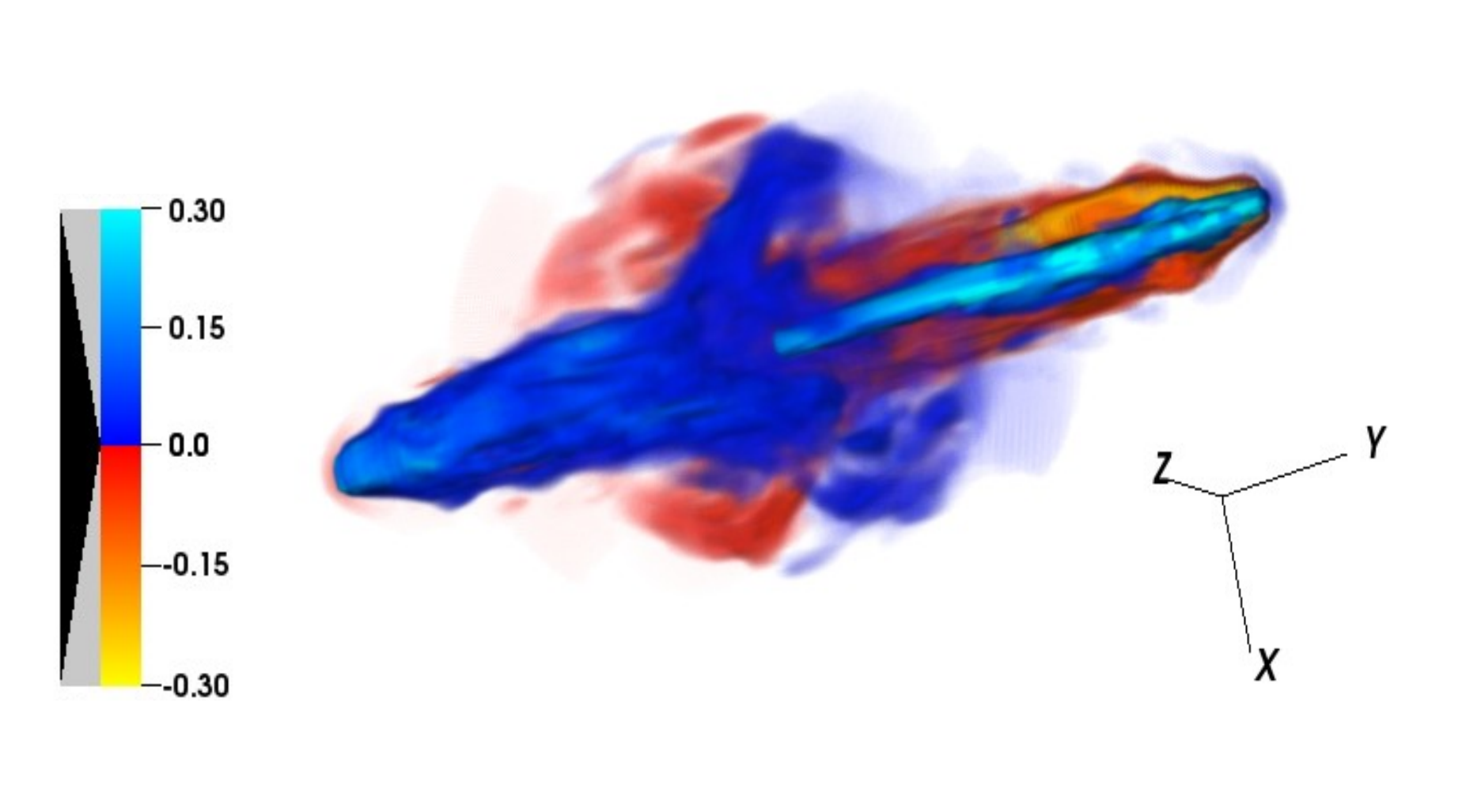}
\caption{\textit{Left:} a 2D slice of our simulated 3D structure (at $x = 0$) representing the density distribution in the $y-z$ plane. The formation of a prominent wing structure along the $z$-axis can be observed here (case: $3Dmaj\_ref$). \textit{Right:} velocity distribution map $(\rm{v}_y)$ of the bi-directional jet, representing the same structure as the \textit{left} one but viewed at an angle to the jet ejection axis. Both of these structures have a dynamical age of 3.91 Myr. The measure of velocity, density and length are represented here with respect to the light speed ($c$), unit density ($\rho_0$ i.e. 1 amu/cc) and unit length ($L_0^{\rm 3D}$ i.e. 4 kpc) respectively.}
\label{Fig:3D}
\end{figure*}

At the onset when the jet head enters into the galactic medium, it results in the formation of a spherical cocoon due to the high-density distribution near the center of the galaxy. As the time progresses, both the jet and counter-jet propagates outwards and exit the high density galactic core region. The existing over-pressured spherical cocoon starts to expand asymmetrically, following the maximum pressure gradient path created by the ambient medium. Elliptical galaxies have higher pressure gradient along the minor axis compared to the major one. Thus, when jets travel along the major axis, the back-flowing material (that contributes to spherical cocoon) expands faster along the minor axis creating a prominent wing structure. In Fig. \ref{Fig:3D} (for the case $3Dmaj\_ref$), we have shown the density distribution (left) viewed along the $x$-axis indicating the formation of a prominent wing structure. On the right, we have shown three dimensional velocity distribution $(\rm v_{y})$ viewed at an angle to the jet ejection axis. This velocity distribution demonstrates the formation of the back-flowing plasma at the jet head and accumulation of it in the wings. The corresponding time of evolution for both these structures is 3.91 Myr.

\begin{figure}
\centering
\includegraphics[width=\columnwidth]{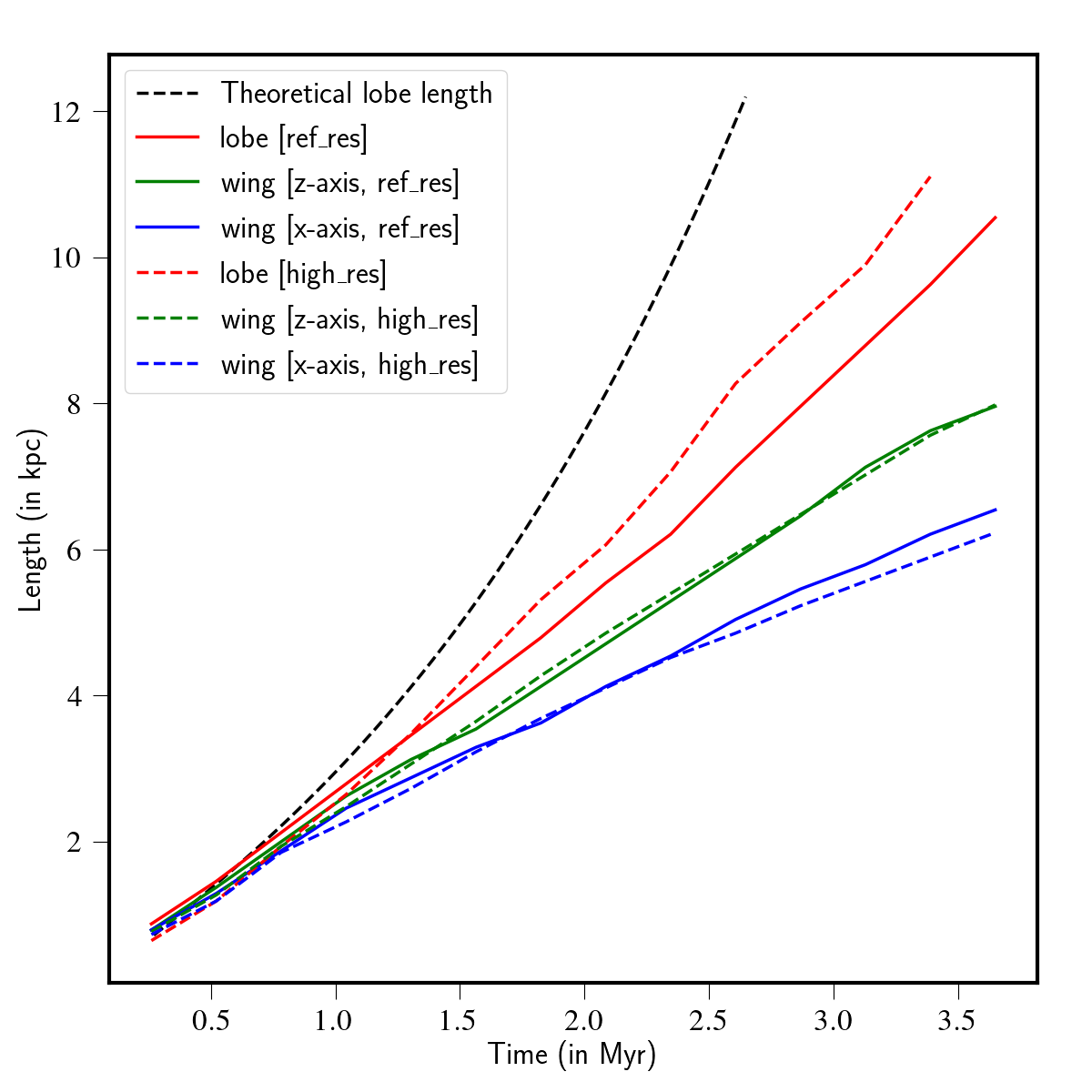}
\caption{A comparison of the theoretical estimation of lobe length with our simulated lobe length obtained for the run $3Dmaj\_ref$ is shown. The time evolution of the wing lengths along the intermediate axis (i.e. along $x$-axis) and the minor axis (i.e. along $z$-axis) along with the comparison with the high resolution 3D run ($3Dmaj\_high$) are also overlaid. The caption ref\_res used here for the run $3Dmaj\_ref$ and high\_res for the run $3Dmaj\_high$.}
\label{Fig:l3D}
\end{figure}

\begin{figure}
\centering
\includegraphics[width=\columnwidth]{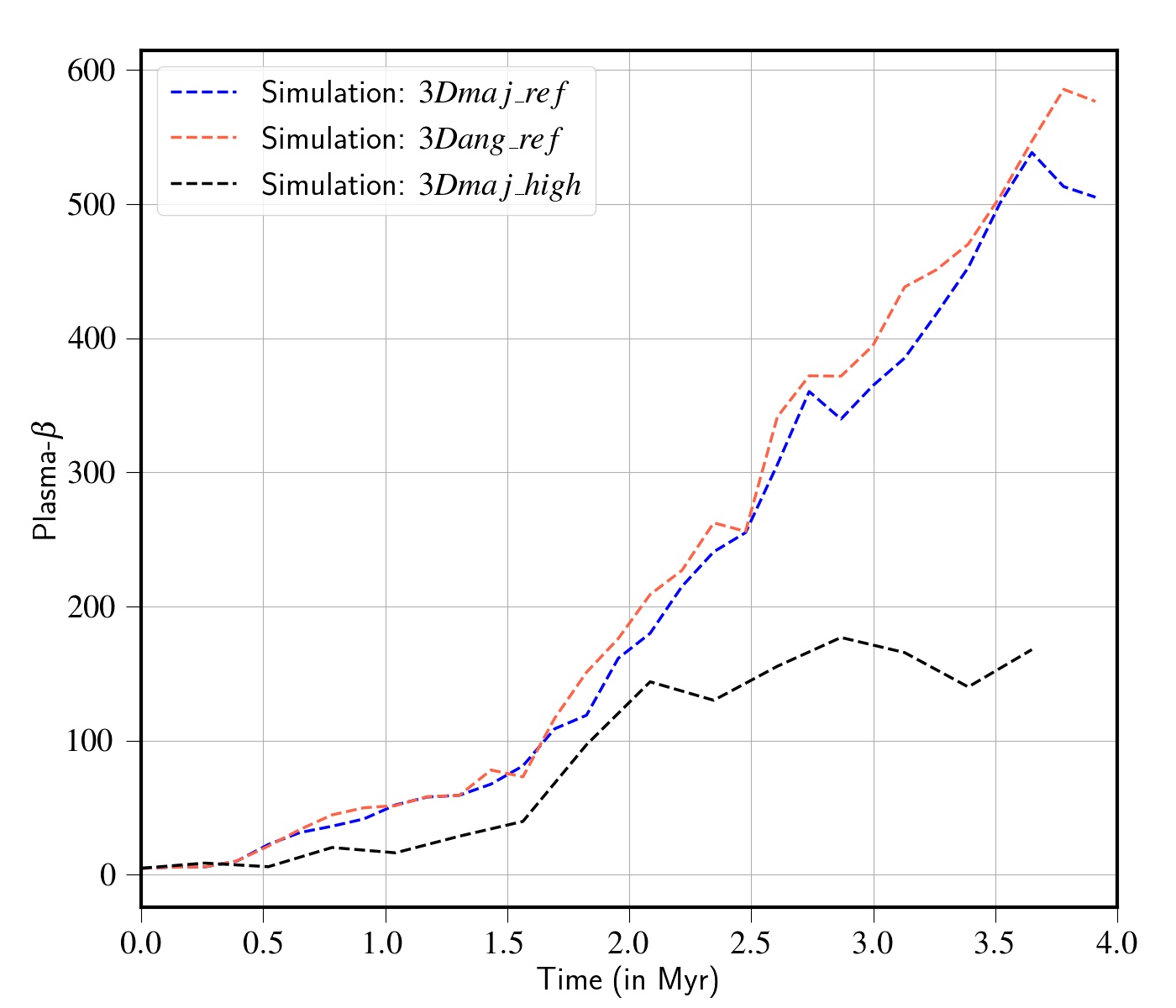}
\caption{Time evolution of plasma-$\beta$ ($\beta_{\rm pl}$) for all of our runs showing a gradual increasing nature throughout. The values of it obtained are much higher than unity indicating the presence of a thermally over-pressured cocoon.}
\label{Fig:pbeta}
\end{figure}

A comparison with the theoretical estimation of the active lobe length with the simulated length is also made for this case and is represented in Fig. \ref{Fig:l3D} where in addition to it, a comparison with our high resolution run is also overlaid. The theoretical estimation of the active lobe length is obtained from the equation as suggested by \citet{Marti1997} for relativistic case i.e. 
\begin{equation} \label{eq:5}
   {\rm v}_h = \frac{\Gamma_j \sqrt{\chi}}{1 + \Gamma_j \sqrt{\chi}} {\rm v}_j;\ {\rm where}\ 
   \chi = \frac{\rho_j h_j}{\rho_a h_a}
\end{equation}
where ${\rm v}_h$, $\Gamma_j$, $\rho_j$, $\rho_a$, $h_j$, $h_a$ are representing jet head speed, jet Lorentz factor, jet density, ambient density and jet and ambient specific enthalpies respectively. In order to determine the simulated active lobe length, we have tracked the maximum pressure region which will occur at the jet head-ambient interaction point \citep{Bromberg2011}. Whereas for the wing length determination, we have tracked the maximum value of kinetic energy along the wing. In Fig. \ref{Fig:l3D}, we can see that the simulated line never crosses the theoretical prediction (1D), as we have extra dimensions here (3D) where the jet can expand, reducing the jet velocity further \citep{Rossi2017}. It is also noticeable that with increasing resolution, the results approach the theoretical prediction. However, the deviation in the wing length and the lobe length for our high and reference resolution cases are $<$ 4\% and 12\% respectively at a time 3.13 Myr. This is equivalent to a lag of $<$ 0.2 kpc and 1.1 kpc for the wing and lobe length, compared to their total size of 5.6 kpc and 9.9 kpc respectively.
The average values of magnetic field (density weighted), we get at the end of the simulations (3.91 Myr) are of the order of 12 $\mu$G. Similarly, for the average pressure of the cocoon, the values we get is of the order of $10^{-9} \  \rm dyn/cm^2$.  
In addition to this, we have also plotted the time evolution of plasma-$\beta$ (the ratio of thermal to magnetic pressure, $\beta_{\rm pl}$ $ = \frac{8\pi P}{B^2}$) in order to understand which form of energy plays the crucial role in shaping these galaxies. The values of $\beta_{\rm pl}$ represented in Fig. \ref{Fig:pbeta} are much higher than unity. This indicates that it is the thermal part that remains dominated in the cocoon over a long time for these galaxies (thermally over-pressured cocoon). Also at the same time, the cocoon is strongly over-pressured with respect to the ambient as well. In this regard, the evolution of $\beta_{\rm pl}$ obtained for the case $3Dmaj\_high$ does not show a steep increase, however an overall increasing nature with $\beta_{\rm pl} \gg 1$ of the curve is obvious.

\subsection{Jet at an angle to major axis} \label{Jet at an angle to major axis}
\begin{figure}
\centering
\includegraphics[width=\columnwidth]{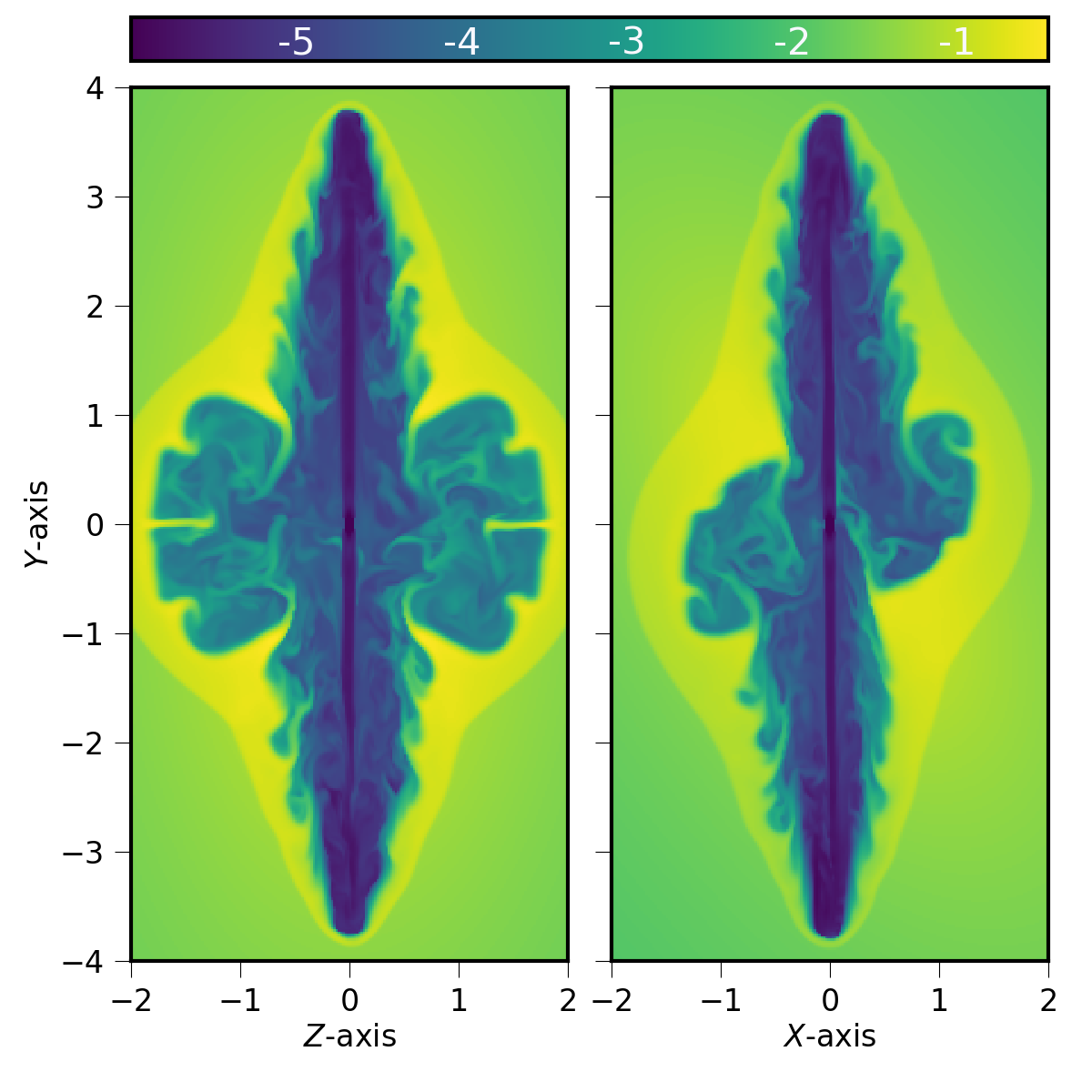}
\caption{Density distribution ($\rm log \ \rho$) of run $3Dang\_ref$ in the plane $y$-$z$ (\textit{left}) and $y$-$x$ (\textit{right}) showing the wing formation along the $z$ and $x$ axis respectively. The asymmetry in wing formation i.e. the bending of back-flowing materials before reaching centre can be seen here along the $x$-axis, whose morphological relevance is discussed in Section \ref{Jet at an angle to major axis}. The dynamical age of this structure is 3.91 Myr. Here lengths are defined with respect to 4 kpc.}
\label{Fig:ang_dyn}
\end{figure}

For the simulation $3Dang\_ref$, where the jet propagates at an angle $30^{\circ}$ to the major axis of the galaxy, still favours the formation of an X-shape morphology. However, as it deviates from the most favourable condition of wing formation, the prominence of the formed structure is less than the run $3Dmaj\_ref$ \citep{Rossi2017}. In Fig. \ref{Fig:ang_dyn}, we have represented the density distribution of the galaxy along the plane $y$-$z$ (left) and $y$-$x$ (right) where the lateral expansion of the cocoon is still in progress. An asymmetry in wing formation i.e. the bending of back-flowing materials before reaching centre can be seen in the $y$-$x$ plane i.e. when viewed along the $z$-axis of the galaxy, where a similar bending of back-flowing plasma has also been observed in several XRGs as well \citep{Cotton2020,Bruni2021}. However, our structure is still in its initial stage of formation (3.91 Myr) and in this regard, it would be interesting to further follow its long term evolution. The lobe length obtained for this case is found to be 24\% higher than the case $3Dmaj\_ref$ at a time 3.65 Myr, whereas no significant difference in wing lengths is observed for both the cases. This indicates the crucial role of thermally over-pressured cocoon in driving the wings beside the leading role played by ambient at its first place. The higher lobe length obtained here is also expected due to the higher propagation speed of the jet \citep{Rossi2017} which is generated due to the faster decay of ambient pressure along the jet propagation direction. The corresponding evolution of plasma-$\beta$ follows a similar trend as $3Dmaj\_ref$ (Fig. \ref{Fig:pbeta}), however the values obtained are slightly higher.

In summary, our 3D relativistic MHD simulations of jet propagation within a tri-axial galactic environment have demonstrated the formation of X-morphology following the Back-flow model as proposed by \citet{Capetti2002}. One of the observed characteristics of such X-morphology is that in most cases wings coincide with the minor axis of the galaxy \citep[e.g.,][]{Saripalli2009,Gillone2016,Joshi2019}. We have performed simulation runs by varying the direction of jet ejection with respect to the major axis of the underlying galaxy and verified such empirical evidence consistent with results obtained by \citet{Rossi2017}. For these radio galaxies, the dominant thermal energy over magnetic energy favours the development of X-shaped morphology and it remains dominated in the cocoon during the evolution time. The back-flowing plasma which were created at the jet head due to the pressure imbalance \citep{Leahy1984}, continuously inflates the cocoon and get accumulated in the wings.

\section{Synthetic Spectral Characteristics} \label{Results 3D: Synthetic spectra and the view angle input}

\subsection{Effect of viewing angle on Surface Brightness Maps}
\begin{figure*}
\centering
\includegraphics[width=17.6cm, height = 10.5 cm]{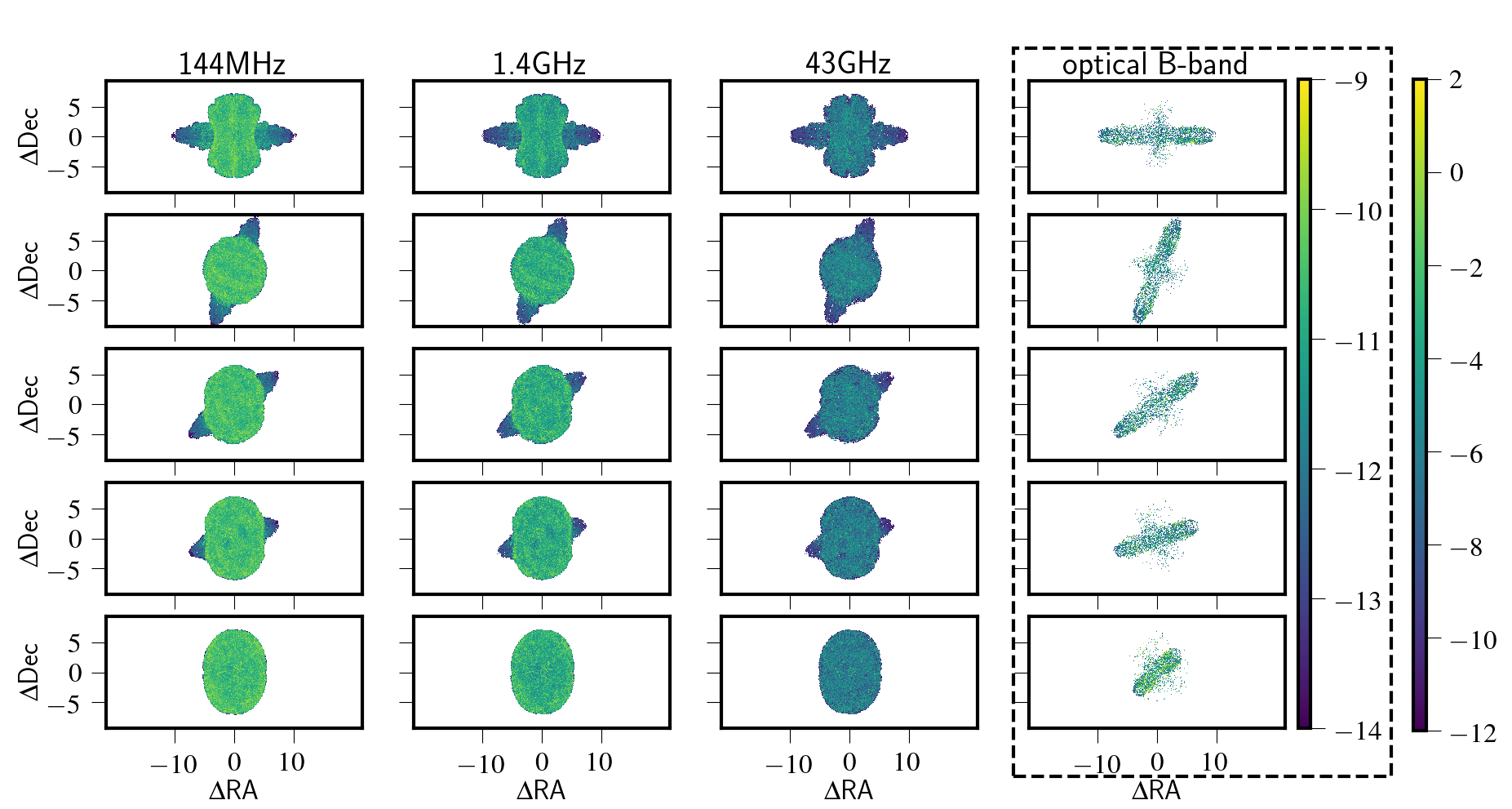}
\caption{Intensity maps ($\rm log$ I in $\rm Jy/arcsec^2$) obtained for the run $3Dmaj\_ref$ are highlighted in this image for five line-of-sight angles i.e. ($\theta = 70^{\circ},\phi = 0^{\circ}$), ($\theta = 20^{\circ},\phi = 70^{\circ}$), ($\theta = 45^{\circ},\phi = 45^{\circ}$), ($\theta = 70^{\circ},\phi = 45^{\circ}$) and ($\theta = 70^{\circ},\phi = 70^{\circ}$) for 1st, 2nd, 3rd, 4th and 5th row respectively (\textit{from top}) at a time 3.78 Myr. We have used a separate color-bar for the optical image to highlight the small scale variation in the intensity distribution. It is evident from figure that the relative extent of the active lobes and the wings depend on the viewing effect. The measure of $\Delta$RA and $\Delta$Dec here are in arcsecond.}
\label{Fig:common_plot_sync_3D}
\end{figure*}

Using the Lagrangian macro-particles injected into the domain, we have also evaluated the intensity distribution of the galaxy. 
The results for case $3Dmaj\_ref$ are shown in Fig. \ref{Fig:common_plot_sync_3D} where in addition to the intensity map, the effect of viewing on the shape is also demonstrated. 
The line of sight angle of visualization from the top row to bottom row in Fig. \ref{Fig:common_plot_sync_3D} are ($\theta = 70^{\circ} ,\phi = 0^{\circ}$), ($\theta = 20^{\circ} ,\phi = 70^{\circ}$), ($\theta = 45^{\circ} ,\phi = 45^{\circ}$), ($\theta = 70^{\circ} ,\phi = 45^{\circ}$) and ($\theta = 70^{\circ} ,\ \phi = 70^{\circ}$) respectively. 
In the figure, each column represents emission maps for different observed frequencies as mentioned on top of the column. A common color-bar for the radio frequencies is shown in the utmost right of the figure. A separate color-bar for the optical (B-band) map (separated with a dashed rectangular block) is used in order to clearly demonstrate the small scale variations of intensity. 
From Fig. \ref{Fig:common_plot_sync_3D}, it is evident that the relative extent of the active lobe and the wing also depends on the viewing angle. Our projected surface brightness maps, synthesized from a single simulation, show similarities with several observed XRGs in terms of appearance.
As suggested by \citet{Hodges-Kluck2011} that the projection effect almost always enhances the wing size, can also be found in our study as well.

We observe that as the viewing angle $\theta$ approaches the intermediate-axis ($x$-axis), the contribution from the wing to the synthetic emission increases in comparison to that obtained from lobe. Similarly, as the value of $\phi$ increases, the contribution from wing to the emission also increases. In the odd case with very high $\theta = 70^{\circ}$ and $\phi=70^{\circ}$, emission in the radio bands from wing over-powers that from the lobe and the observed structure is rather a blob and can not be categorized as X-shape. 
The projected lobe length to wing length ratio i.e. $\mathcal{L}_{\rm lobe}/\mathcal{L}_{\rm wing}$ for the case ($70^{\circ},70^{\circ}$) is 0.7, whereas, it is $\approx 1$ for ($70^{\circ},\ 45^{\circ}$), stipulating a population that possess equal projected wing and lobe length. 
The values of $\mathcal{L}_{\rm lobe}/\mathcal{L}_{\rm wing}$ for other cases and viewing angles are listed in Table \ref{Tab:delta_alpha_dist} which are typically $> 1$.

In summary, we find that synthetic emission maps produced from simulations with back flow can span the full range of $\mathcal{L}_{\rm lobe}/\mathcal{L}_{\rm wing}$ with different viewing angles. This result is consistent with the findings of \citet{Hodges-Kluck2011,Joshi2019}.  

\subsection{Spectral Maps and Energy Distribution} \label{Spectral Maps and Energy Distribution}
\begin{figure}
\centering
\includegraphics[width=\columnwidth,height=6cm]{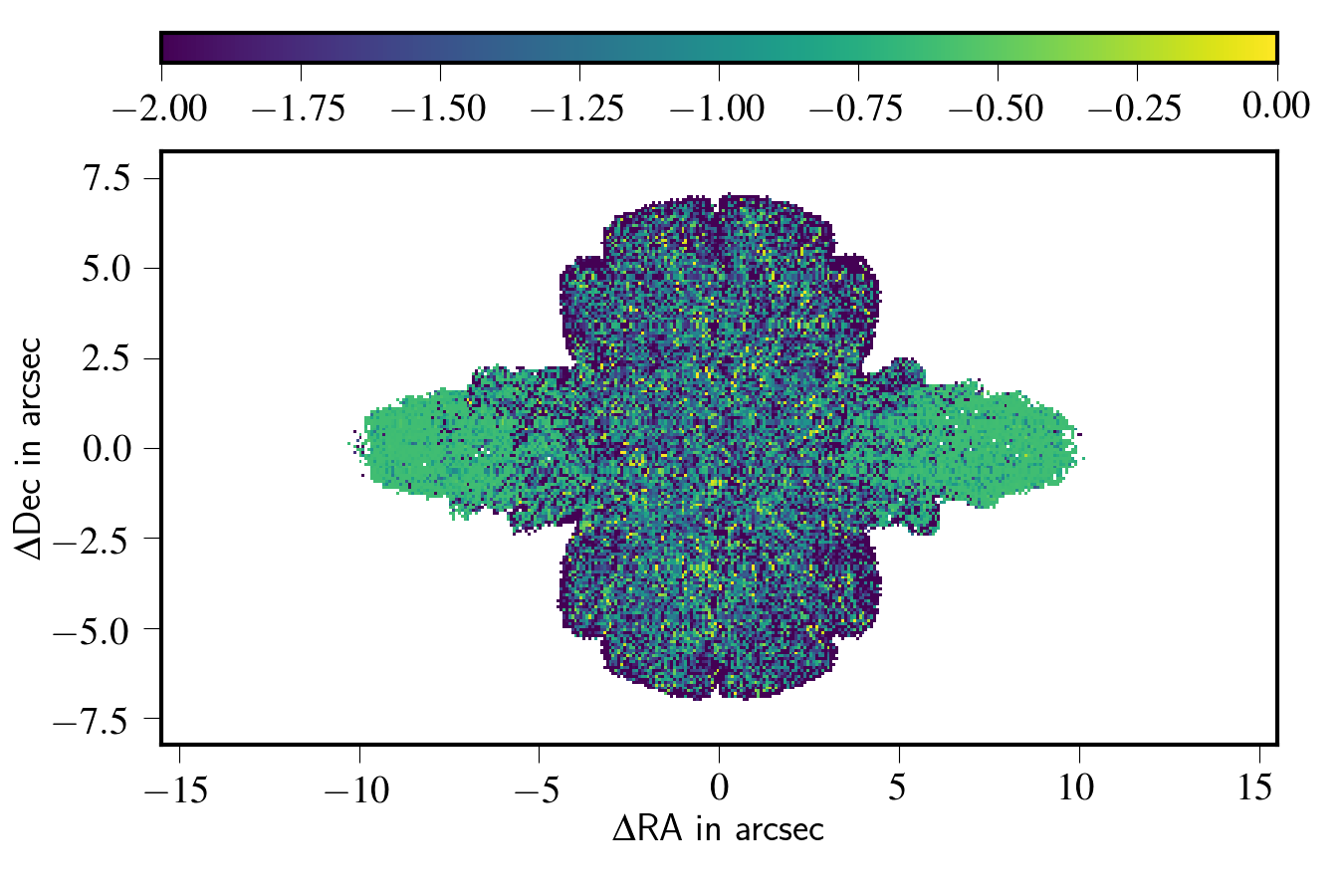}
\caption{The spectral index distribution ($\alpha$) of the radio structure obtained for the run $3Dmaj\_ref$ is shown. Here wings are showing steep spectral index compared to the active lobes due to the accumulation of older particles. However, the presence of several flatter $\alpha$-patches are also noticeable (especially in the wings) that are generated due to re-energized particles at shocks. The frequencies used here to produce this map are 240 MHz and 5 GHz.}
\label{Fig:spectral_index}
\end{figure}

From the power-law dependence of the flux density $S_{\nu}\propto\nu^{\alpha}$, the spectral index $\alpha$ can be obtained as 
\begin{equation}
\label{eq:specindx}
    \alpha = \log_{10}(S_{\nu1}/S_{\nu2})/\log_{10}(\nu1/\nu2)
\end{equation}
Its distribution for the $3Dmaj\_ref$ case with viewing angle ($\theta = 70^{\circ} ,\phi = 0^{\circ}$) and for frequencies $\nu1 =$ 240 MHz and $\nu2 =$ 5 GHz is shown in Fig. \ref{Fig:spectral_index}. 
The intensity weighted average of $\alpha$ of the whole structure is evaluated using the equation below
\begin{equation} \label{eq9}
    \alpha_{\rm av} = \frac{\int I_{\nu}(x,y)\alpha(x,y) dx dy}{\int I_{\nu}(x,y) dx dy} 
\end{equation}
and the value obtained is -1.16. 
From the spectral map distribution, it is evident that steeper spectra is primarily associated with the wings (dark blue color) while the lobes show a rather flat spectral signatures (light green color). 
Considering the wings, we also observe that at significant places, the spectral index values becomes flat (localized green spots in the wing structure). This indicates possibility of particle energisation due to localised shocks within the wings. We have further analysed particle distribution both in the wings and the lobes separately and is discussed in Section \ref{Results 3D: Particle Properties}.

\begin{figure*}
\centering
\includegraphics[width=2\columnwidth]{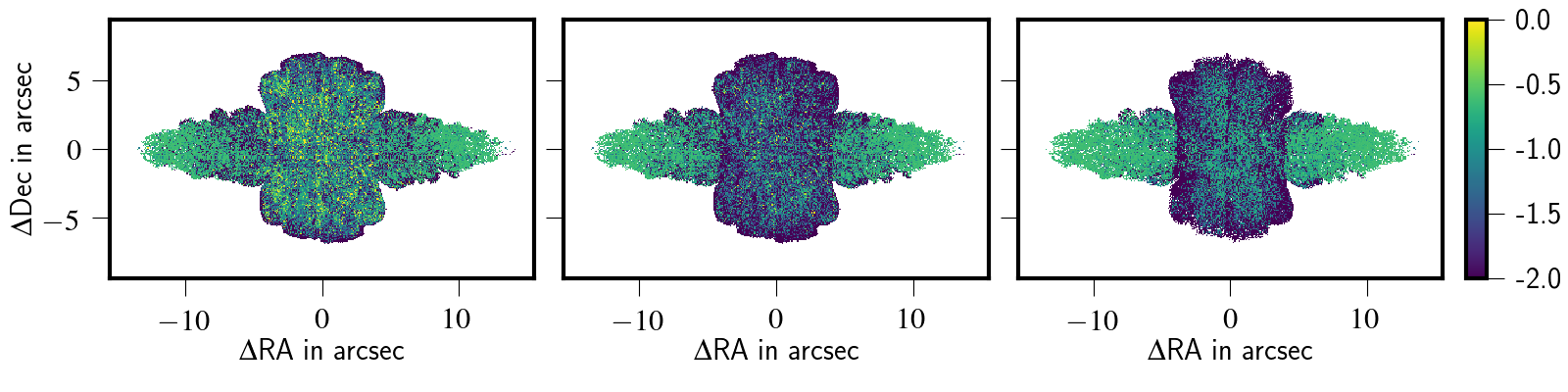}
\caption{Distribution of spectral index values for different frequency choices for the case $3Dang\_ref$ is shown here for a line of sight angle of ($70^{\circ},0^{\circ}$). The frequencies used to generate these maps are 0.240-0.610 GHz (\textit{left}), 0.240-5.0 GHz (\textit{middle}) and 5.0-43.0 GHz (\textit{right}) respectively. The steepening of $\alpha$-values with increasing frequency choices in the wings is expected due to synchrotron cooling. The presence of flatter $\alpha$-patches (particularly in the wings) are prominent if the chosen frequencies are in the low frequency regime (indication of particle re-energization).}
\label{Fig:alpha_freq}
\end{figure*}

\begin{figure}
\centering
\includegraphics[width=\columnwidth,height=6.2cm]{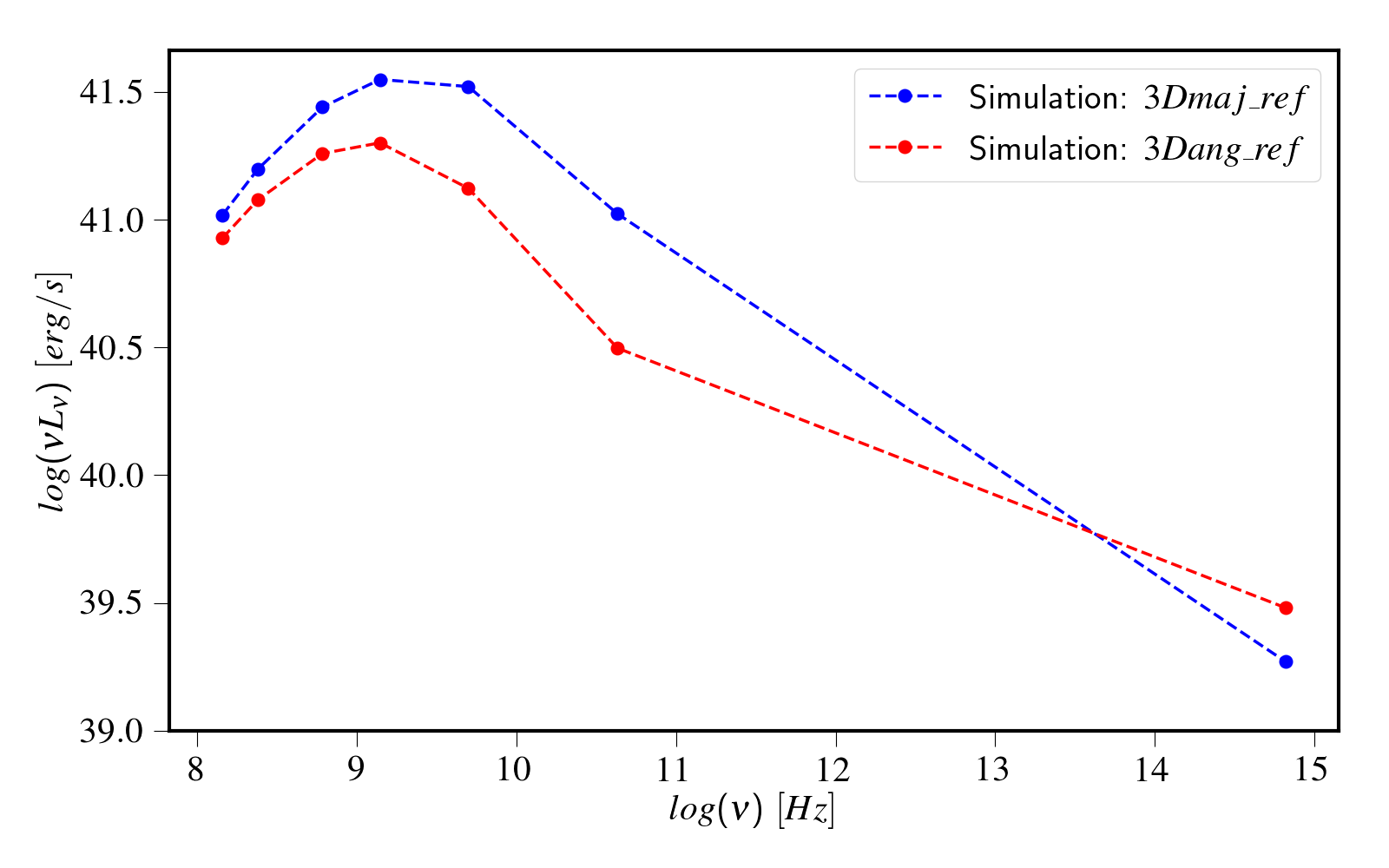}
\caption{The spectral energy distribution (SED) ($\nu L_{\nu}$: $L_{\nu}$ is specific luminosity) of our simulated XRGs in seven synchrotron frequencies are represented here. A low frequency hump, peaking at 1.4 GHz is prominent in the image for both our 3D cases.}
\label{Fig:SED}
\end{figure}

The distribution of spectral index also depends on the choice of frequencies and therefore we have created spectral maps with multiple frequency pairs i.e. for ($\nu1$, $\nu2$) as (0.240, 0.610) GHz, (0.240, 5.0) GHz and (5.0, 43.0) GHz (see Eq.~\ref{eq:specindx}).
In this way, we are probing two ends of a power law spectrum that has a break near 5 GHz observed in our case. The $\alpha$-maps with the above frequency choices are shown in Fig. \ref{Fig:alpha_freq} for the case $3Dang\_ref$ with line of sight angle ($70^{\circ},0^{\circ}$). The maps indicate that spectral index is flat within the lobes for all chosen frequency pairs. Whereas, wings demonstrate a much steeper value of $\alpha$, particularly for higher frequency choices (see middle and right panel of Fig.~\ref{Fig:alpha_freq}).
This corroborates the fact that synchrotron cooling time is small for high energy particles and verifies that older particles (those injected at early times in the simulation) have considerably cooled due to radiative processes and accumulated in the wings. It should be noted that even for the $\alpha$-map with 5 and 43 GHz (right panel of Fig.~\ref{Fig:alpha_freq}), one can observe local patches in the wings that corresponds to flat spectra with $\alpha \sim -0.7$ primarily arising due to shocks.  
The associated spectral energy distribution ($\nu L_{\nu}$ with $L_{\nu}$ as the specific luminosity) is obtained for seven synchrotron frequencies and presented in Fig. \ref{Fig:SED}. We observe a low frequency hump peaking at 1.4 GHz for both our 3D cases with a peak value of $3 \times 10^{41}$ erg/sec. Also, the luminosity values show a similar declining trend with higher frequencies as expected due to synchrotron cooling.\\


\begin{figure*}
\centering
\includegraphics[width=2\columnwidth]{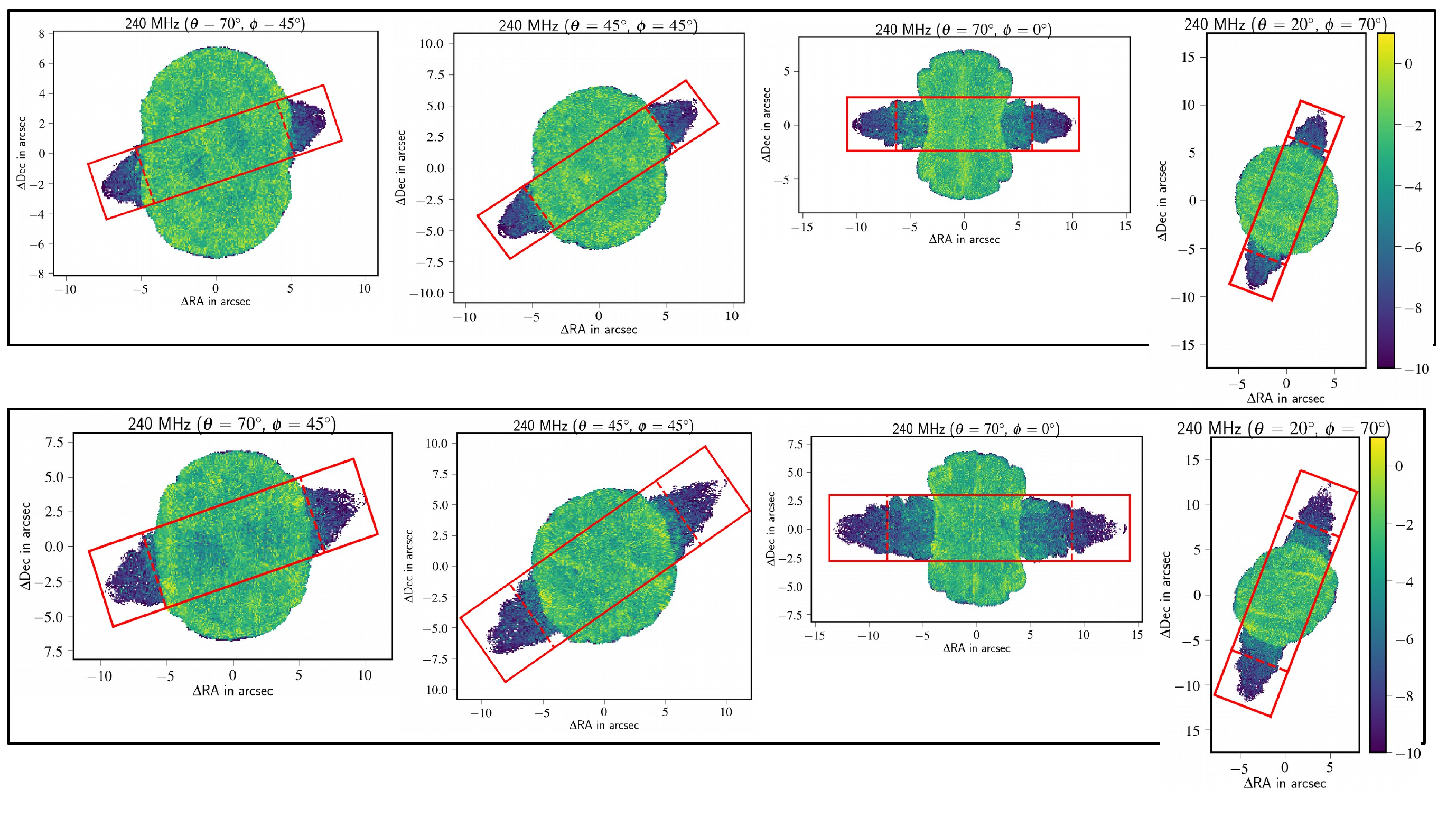}
\caption{\textit{Top}: Intensity maps, obtained for four line of sight angles for the case $3Dmaj\_ref$, \textit{bottom}: intensity maps, obtained for the same los angles for case $3Dang\_ref$. We have marked the jet region with red rectangle (solid line), outside of which, it is the wing region. The rectangles drawn in the above images ensure that the jet remains well fit within that region which is verified using our optical-B band images. The area enclosed by the dashed line indicates the tip of the jet (20\% of the projected jet length) accounted to avoid contamination while measuring $\Delta\alpha_{\nu1}^{\nu2}$. The corresponding color-bar represents the 240 MHz intensity measures ($\rm log\ I$) in $\rm Jy/arcsec^2$.}
\label{Fig:los_alpha}
\end{figure*}

We have further evaluated $\Delta \alpha$, the difference between the averaged spectral index of the active lobes ($\alpha^{\rm lobe}_{\rm av}$) and the wings ($\alpha^{\rm wing}_{\rm av}$), for different viewing angles and different pairs of frequencies. We denote this quantity as $\Delta \alpha ^{\nu2}_{\nu1}$ and compute the same using 
\begin{equation}
\label{eq:deltaalpha}
    \Delta \alpha_{\nu1}^{\nu2} = \alpha^{\rm lobe}_{\rm av} - \alpha^{\rm wing}_{\rm av}
\end{equation}
We have separated out the region of the lobe from the the wing as indicated by red rectangle with solid line in Fig. \ref{Fig:los_alpha}. The rectangle is drawn ensuring that the jet structure obtained from optical B-band images remains well within its sides. The contribution from the overlapping region is eliminated by identifying the lobes by only considering 20\% of the projected jet length as shown by the red dashed lines. The region outside the solid rectangle is identified as wing. 
The intensity averaged spectral index is finally estimated using Eq. (\ref{eq9}) for each of these regions. It is clear from the expression of $\Delta \alpha _{\nu1}^{\nu2}$ (see Eq.~\ref{eq:deltaalpha}) that a negative ($-$ve) value would indicate flatter spectral index in wings as compared to lobes. While, a positive ($+$ve) $\Delta \alpha _{\nu1}^{\nu2}$ would imply that lobes have flatter spectral index compared to the wings.

The values of $\Delta \alpha_{\nu1}^{\nu2}$, obtained for frequencies $\nu1=240$ MHz and $\nu2=610$ MHz are highlighted in Table \ref{Tab:delta_alpha_dist} for both our reference runs. We find that its value can be either $+$ve or $-$ve depending on the viewing angle. The measure of $\Delta \alpha ^{\nu2}_{\nu1}$ is also sensitive to the choices of frequencies. 
We have obtained its values with different sets of $\nu1$ and $\nu2$ which are listed in Table \ref{Tab:delta_alpha_dist2}. The choices are made in such a way that we can probe $\Delta\alpha_{\nu1}^{\nu2}$ in the two regime of a broken power law spectrum obtained for our radio galaxy (spectral break around 5 GHz). We see that $\Delta\alpha_{\nu1}^{\nu2}$ values are mostly $+$ve for higher frequencies and also have higher values, as is expected of older cooling electrons in the wings with steeper spectral indices. 
Negative values of $\Delta \alpha ^{\nu2}_{\nu1}$ are seen for the pairs of frequencies that are in the rising phase of SED (up to the peak at 1.4 GHz). 
An exception is where $\Delta \alpha^{\rm 43 GHz}_{\rm 5 GHz}$ is $-$ve, regardless the fact that both the lobe and wing spectral index ($\alpha_{\rm av}^{\rm lobe}$, $\alpha_{\rm av}^{\rm wing}$) steepens for these frequencies. This is the case with highest projected wing size with a small tip of jet visible in the intensity map. This further hints that the projected size of lobes and wings may play an integral role in influencing the sign of $\Delta \alpha ^{\nu2}_{\nu1}$. Thus, viewing angle also plays a crucial role in determining the value of $\Delta\alpha^{\nu2}_{\nu1}$ as shown in Tables~\ref{Tab:delta_alpha_dist} and \ref{Tab:delta_alpha_dist2}. In case of our reference run $3Dmaj\_ref$, the line of sight angles ($70^{\circ}, 0^{\circ}$) and ($70^{\circ}, 20^{\circ}$) provide the highest projected lobe length to wing length ratio ($\geq 1.5$) and correspondingly have $-$ve values of $\Delta \alpha^{610\ \rm MHz}_{240\ \rm MHz}$. 
Whereas, the other two viewing angles have larger wing lengths and so values of $\Delta\alpha^{610\ \rm MHz}_{240\ \rm MHz}$ obtained are $+$ve. 
For the case $3Dang\_ref$, the obtained lengths of lobes are naturally higher due to the higher propagation speed of the jet, as also discussed in Section \ref{Jet at an angle to major axis}. 
In this case, all the values obtained for $\Delta \alpha_{\nu1}^{\nu2}$ are $-$ve for 240 MHz ($\nu1$) and 610 MHz ($\nu2$). However, this is not a general trend observed for all the frequency choices. We have also revisited our calculation (i.e. measure of $\Delta \alpha_{\nu1}^{\nu2}$) following the same process as highlighted above but varying the tip length of the jet from 15\% to 25\%. We observed that changes in ways to distinguish lobes and wings only affects the quantitative values, but the overall qualitative picture presented here still holds true. This is expected as with different line of sight choices and box sizes, we are considering or ignoring regions where effects from nonlinear micro-physical processes are also in play. As a result, a wide range of $\Delta \alpha_{\nu1}^{\nu2}$ can be anticipated from a single source based on the way we measure them. In case of a bigger box shape, the effect of contamination (between lobe and wing) shall be imparted on the $\Delta \alpha_{\nu1}^{\nu2}$ measurements.

From an observation study of 28 XRGs, \cite{Lal2019} have demonstrated that the values of $\Delta \alpha_{240\,\rm MHz}^{610\,\rm MHz}$ approximately varies between -0.7 to +0.6. 
Based on such variation, their study did not find any significant difference between the spectral behaviour of wings and lobes and suggested a possible role of dual AGN in formation of XRGs. From our analysis, we have showed that $\Delta \alpha_{\nu1}^{\nu2}$ distribution has a strong dependence on frequency choices, especially which part of SED they are in. In this regard, one should note that the peak frequency in SEDs of several observed XRGs has been found to vary in a wider range starting from MHz to GHz scale \citep{Lal2007,Ignesti2020}. 
The distribution also shows variation depending on the viewing angles. As the size and shape of the structure changes based on line of sight visualization angles, so does the distribution of $\alpha$ which in turn affects the $\alpha_{\rm av}$ evaluation and hence the $\Delta \alpha_{\nu1}^{\nu2}$. The diffusive shocks, especially, complicates it even more by producing flatter-$\alpha$ patches. Further, shock acceleration is a continuous process and so is the radiative cooling processes. Hence, such a measure would critically depend on the evolution stage of the galaxy. Our current models are limited to simulations focusing on the early phase of XRGs. Although, the mechanism through which the shocks are generated are a continuous process, and it is expected that cocoons and wings of larger scale XRGs will also be turbulent as the backflow settles, whether such mechanisms can operate at a later stage of XRGs, and if so, with what efficiency, needs to be tested with long term simulations. Here, we have only intended to highlight the critical dependencies of $\Delta\alpha_{\nu1}^{\nu2}$ on numerous factors that may be in operation in different sources.

In summary, the quantity $\Delta \alpha _{\nu1}^{\nu2}$ has a non-linear dependence on several parameters and have a rather degenerate behaviour and therefore its value may not be sufficient to constrain the formation process of X-shaped radio galaxies. One may find similar distribution of this parameter for the choices of other competing formation models as well, as the shock re-acceleration could also be relevant in these formation scenarios. This existence of degeneracy implies that a measure of difference in spectral index has to be supplemented with other observable to clearly identify the formation mechanism of XRGs. 

\begin{table}
\caption{This table shows the $\Delta \alpha_{\nu1}^{\nu2}$ values that are obtained for frequencies 240 MHz and 610 MHz for our 3D cases, for the four line of sight angles as mentioned earlier. The 4th column represents the projected lobe to wing length ratios for each case. A distribution of $\Delta\alpha_{\rm 240\ MHz}^{\rm 610\ MHz}$ values both in the $+$ve and $-$ve side is particularly noticeable here which has been obtained based on the Back-flow model.}
\begin{center}
\begin{tabular}{ c c c c } 
 \hline
 \vspace{0.1cm}
 Case&($\theta,\ \phi$)&$\Delta \alpha^{610\  \rm{MHz}}_{240\ \rm{MHz}}$&$\mathcal{L}_{\rm lobe}/\mathcal{L}_{\rm wing}$\\
 \hline
 \vspace{0.1cm}
 $3Dmaj\_ref$&$70^{\circ},\ 0^{\circ}$&$-0.294$&1.5\\
 \vspace{0.1cm}
 &$20^{\circ},\ 70^{\circ}$&$-0.200$&1.9\\
 \vspace{0.1cm}
 &$45^{\circ},\ 45^{\circ}$&$+0.092$&1.4\\
 \vspace{0.1cm}
 &$70^{\circ},\ 45^{\circ}$&$+0.156$&1.1\\
 \hline
 $3Dang\_ref$&$70^{\circ},\ 0^{\circ}$&$-0.562$&2.0\\
 \vspace{0.1cm}
 &$20^{\circ},\ 70^{\circ}$&$-0.068$&2.2\\
 \vspace{0.1cm}
 &$45^{\circ},\ 45^{\circ}$&$-0.162$&1.9\\
 \vspace{0.1cm}
 &$70^{\circ},\ 45^{\circ}$&$-0.065$&1.5\\
 \hline
\end{tabular}
\label{Tab:delta_alpha_dist}
\end{center}
\end{table}
\begin{table}
\caption{Here, we have represented the $\Delta\alpha_{\nu1}^{\nu2}$ values, obtained for different choices of frequencies for the case $3Dmaj\_ref$ (\textit{top}) and $3Dang\_ref$ (\textit{bottom}) for four viewing angles. The distribution of $\Delta\alpha_{\nu1}^{\nu2}$ is found to vary based on both the choices of frequencies and view angles (see Section \ref{Results 3D: Synthetic spectra and the view angle input} for details).}
\begin{center}
\begin{tabular}{ c c c c c } 
 \hline
 \vspace{0.1cm}
 ($\theta,\ \phi$)&$\Delta \alpha^{240\  \rm{MHz}}_{144\ \rm{MHz}}$&$\Delta \alpha^{1.4\  \rm{GHz}}_{240\ \rm{MHz}}$&$\Delta \alpha^{5\  \rm{GHz}}_{240\ \rm{MHz}}$&$\Delta \alpha^{43\  \rm{GHz}}_{5\ \rm{GHz}}$\\
 \hline
 \vspace{0.1cm}
 $70^{\circ},\ 0^{\circ}$&$-0.486$&$-0.029$&$+0.522$&$+1.974$\\
 \vspace{0.1cm}
 $20^{\circ},\ 70^{\circ}$&$-0.269$&$-0.082$&$+0.271$&$+2.027$\\
 \vspace{0.1cm}
 $45^{\circ},\ 45^{\circ}$&$-0.102$&$+0.275$&$+0.597$&$+0.560$\\
 \vspace{0.1cm}
 $70^{\circ},\ 45^{\circ}$&$+0.133$&$+0.182$&$+0.260$&$-0.503$\\
 \hline
 $70^{\circ},\ 0^{\circ}$&$-0.869$&$-0.303$&$+0.160$&$+1.840$\\
 \vspace{0.1cm}
 $20^{\circ},\ 70^{\circ}$&$-0.455$&$+0.282$&$+0.923$&$+3.187$\\
 \vspace{0.1cm}
 $45^{\circ},\ 45^{\circ}$&$-0.474$&$+0.158$&$+0.702$&$+1.896$\\
 \vspace{0.1cm}
 $70^{\circ},\ 45^{\circ}$&$-0.317$&$+0.126$&$+0.418$&$+1.339$\\
 \hline
\end{tabular}
\label{Tab:delta_alpha_dist2}
\end{center}
\end{table}

\section{Impact of Shock Acceleration} \label{Results 3D: Particle Properties} 
\subsection{On Maximum Particle Energy}
\begin{figure}
\centering
\includegraphics[width=\columnwidth]{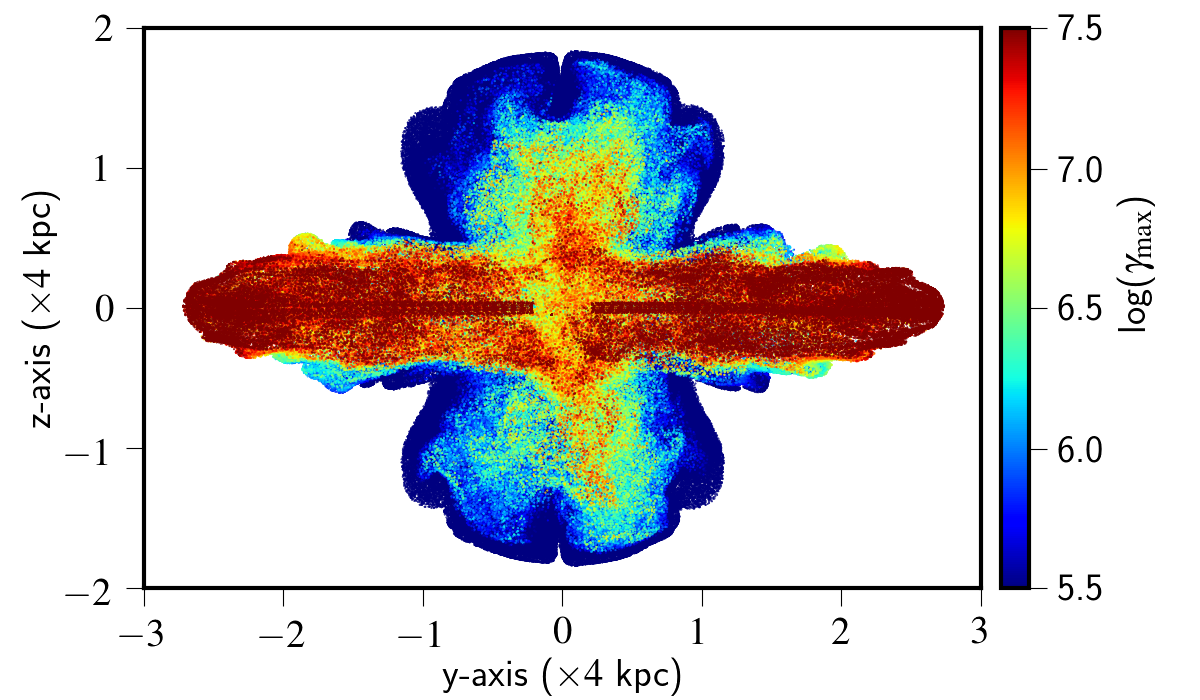}
\caption{Distribution of maximum Lorentz factors ($\gamma_{\rm max}$) of non-thermal electrons of the radio structure at a time 3.78 Myr. This projected distribution (projected on $y$-$z$ plane) provides evidence of re-energization activities in the wings due to shock acceleration. For visualization, we put limit on the colorbar.}
\label{Fig:gamma_m}
\end{figure}

Our simulations have shown that older particles tend to accumulate in the wings that are adiabatically expanding. In this case, one would expect the wing over time to have a rather steep spectral distribution due to losses incurred via radiative and adiabatic processes. Instead, we observe that there exist significant localised flat $\alpha$-patches in the wings at a time when the structure has evolved for 3.78 Myr. 
This indicates that the wing structure of the galaxy is not a passively growing feature, rather it evolves continuously under the influence of particle re-energization activities due to shocks. 
To further validate this fact, we show the maximum Lorentz factor ($\gamma_{\rm max}$) distribution of the non-thermal electrons at a time 3.78 Myr in Fig. \ref{Fig:gamma_m}.
We expect that the initial $\gamma_{\rm max}$ value ($10^{10}$) of the particle distribution will eventually reduce with time due to the presence of radiative and adiabatic losses that we have considered in our simulations. 
However in presence of shocks, at the shock acceleration sites, particle spectra will be flat with an increased $\gamma_{\rm max}$ than that expected from a particle that has never been shock accelerated \citep{borse2021, Mukherjee2021}. Due to the accumulation of older particles in the wings, the $\gamma_{\rm max}$ distribution generally shows somewhat lower values than the lobes. However, the presence of several high $\gamma_{\rm max}$ regions observed in the wings indicates the ongoing re-energizing activities. Here the possible origin of shocks can be attributed to on-going super-sonic turbulence in the wing due to the diverted back flow from the lobes.
To show the correlation between this $\gamma_{\rm max}$ distribution and the obtained $\alpha$-map, we have compared Fig. \ref{Fig:gamma_m} with Fig. \ref{Fig:spectral_index} which we obtained from the same simulation ($3Dmaj\_ref$) and at the same time (3.78 Myr), representing a similar morphology. Considering the active lobe of the structure, the particles get energized in the terminal shocks and begin to cool as they back-flow towards the centre without much re-acceleration event. Some particles even get trapped in the jet side edges, and evolve under the dominant cooling processes reducing their $\gamma_{\rm max}$ values significantly. As a consequence, we see near the jet leading edges the spectral index values become flat, and it is steepening in the side edges of the jet. Also, as we move towards the centre from the jet head, the $\alpha$ values gradually becomes steep. In the secondary lobes however, we see, in a typically steeper $\alpha$-distribution (owing to the accumulation of cooled particles) flatter $\alpha$ patches are in formation governing by the shock accelerations whose strength determines the $\alpha$ values.

\begin{figure}
\centering
\includegraphics[width=\columnwidth]{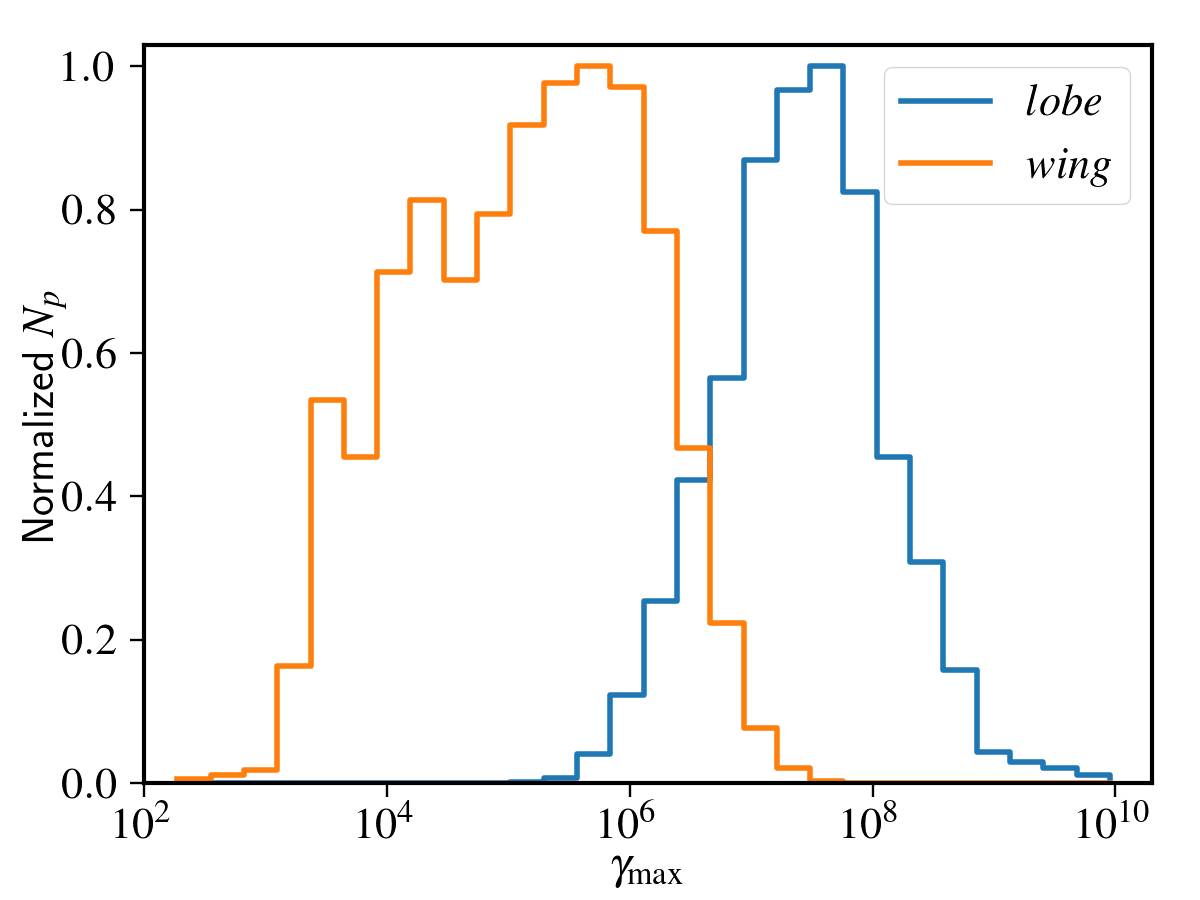}
\caption{Histogram showing the maximum Lorentz factor distribution ($\gamma_{\rm max}$) of lobe and wing particles (considered  without contamination) obtained at a time 3.78 Myr. The vertical axis represents normalized number of particles ($N_p$). We see a lower $\gamma_{\rm max}$ peak for wing distribution in comparison to the lobe that peaks near $10^{7.7}$. The wing distribution also shows a spread in values with a high $\gamma_{\rm max}$ tail (extending up to $\sim 10^{7.7}$) providing evidence of re-energization activities in the wings due to diffusive shocks.}
\label{Fig:gmm_hist}
\end{figure}

To further quantify the distribution shown in Fig. \ref{Fig:gamma_m}, we have separated out the lobe and wing particles following the process highlighted in Section \ref{Spectral Maps and Energy Distribution} (particles in the overlapping region are excluded). We then plotted histogram of $\gamma_{\rm max}$ values of these particles which is shown in Fig. \ref{Fig:gmm_hist}. The histogram confirms the fact discussed above with wing particles peaking at $\sim 10^{5.5}$ in comparison to the lobe particles that peak at $\sim 10^{7.7}$. For the lobe distribution, we see few particles having $\gamma_{\rm max}$ values of $10^{10}$, indicating recently injected particles that still carry the injection spectra. Whereas, most of the lobe particles have started to show cooling effects, resulting in a peak near $10^{7.7}$. For the wing particles, we see broadening of the distribution with the appearance of a high $\gamma_{\rm max}$ tail (extending up to $10^{7.7}$) indicating a mixing of particles that have different evolution histories \citep[see][]{Mukherjee2021}. We have further elaborated this in Section \ref{On Equi-partition Approximation}. The high $\gamma_{\rm max}$ tail of the wing distribution in Fig. \ref{Fig:gmm_hist} is particularly interesting as they carry the signatures of particle re-acceleration process that are generated due to diffusive shocks in the wings.

\begin{figure}
\centering
\includegraphics[width=\columnwidth]{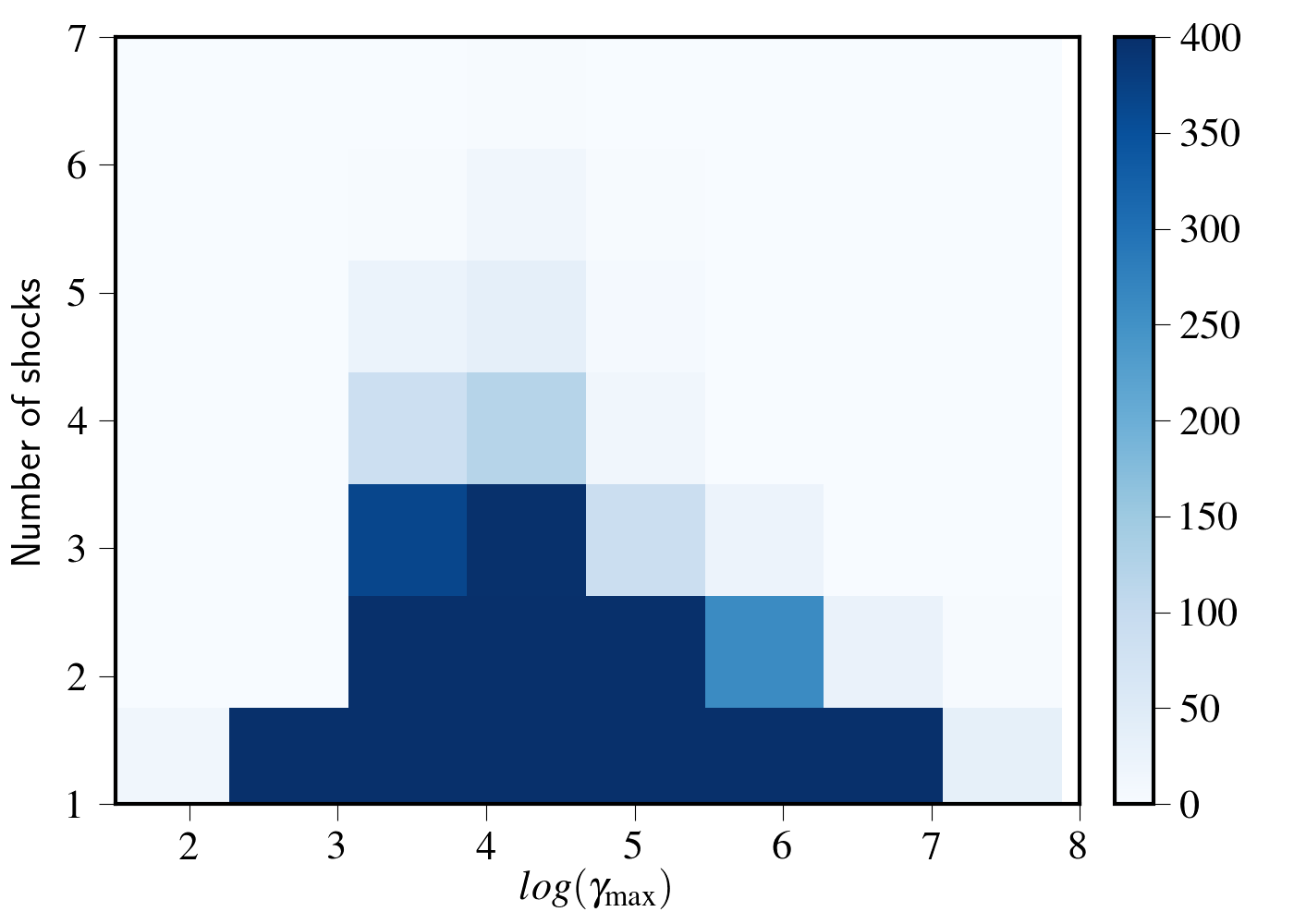}
\caption{2D PDF of number of shocks that the wing particles have encountered vs their corresponding $\gamma_{\rm max}$ values. We see that significant wing particles have experienced greater number of shocks than the lobes ($\leq 2$), providing hint of ongoing re-energization activities in the wings. The corresponding colorbar represents the number of particles that are distributed in the bins, where we put limit for visualization.}
\label{Fig:shk}
\end{figure}

To quantify the impact of shock acceleration, we have analysed the number of shocks that the particles have encountered (in the lobe and wings, obtained without contamination) with their corresponding $\gamma_{\rm max}$ values at time 3.78 Myr. Since the streamlines of some particles have not crossed a strong shock while being advected with the fluid, the particles involved in this analysis represent a sub-sample of particles involves in Fig. \ref{Fig:gmm_hist}. Such a distribution showcases the history of particle since its injection into the numerical domain. 
This distribution is represented via a 2D-histogram in Fig. \ref{Fig:shk} where we show only the particles that are in the wings.
We found that the lobe particles have higher $\gamma_{\rm max}$ values and the number of shocks they have encountered is less ($\leq 2$). Whereas, significant number of wing particles have undergone multiple shocks ($\geq 2$) in their lifetime and show a $\gamma_{\rm max}$ distribution peaking at $\sim 10^{4.5}$ as can be seen in Fig. \ref{Fig:shk}. This again affirms the role of shock acceleration in re-energizing particles in the wings which plays a crucial role in keeping the wing structure active during the evolution time.

\begin{figure}
\centering
\includegraphics[width=\columnwidth]{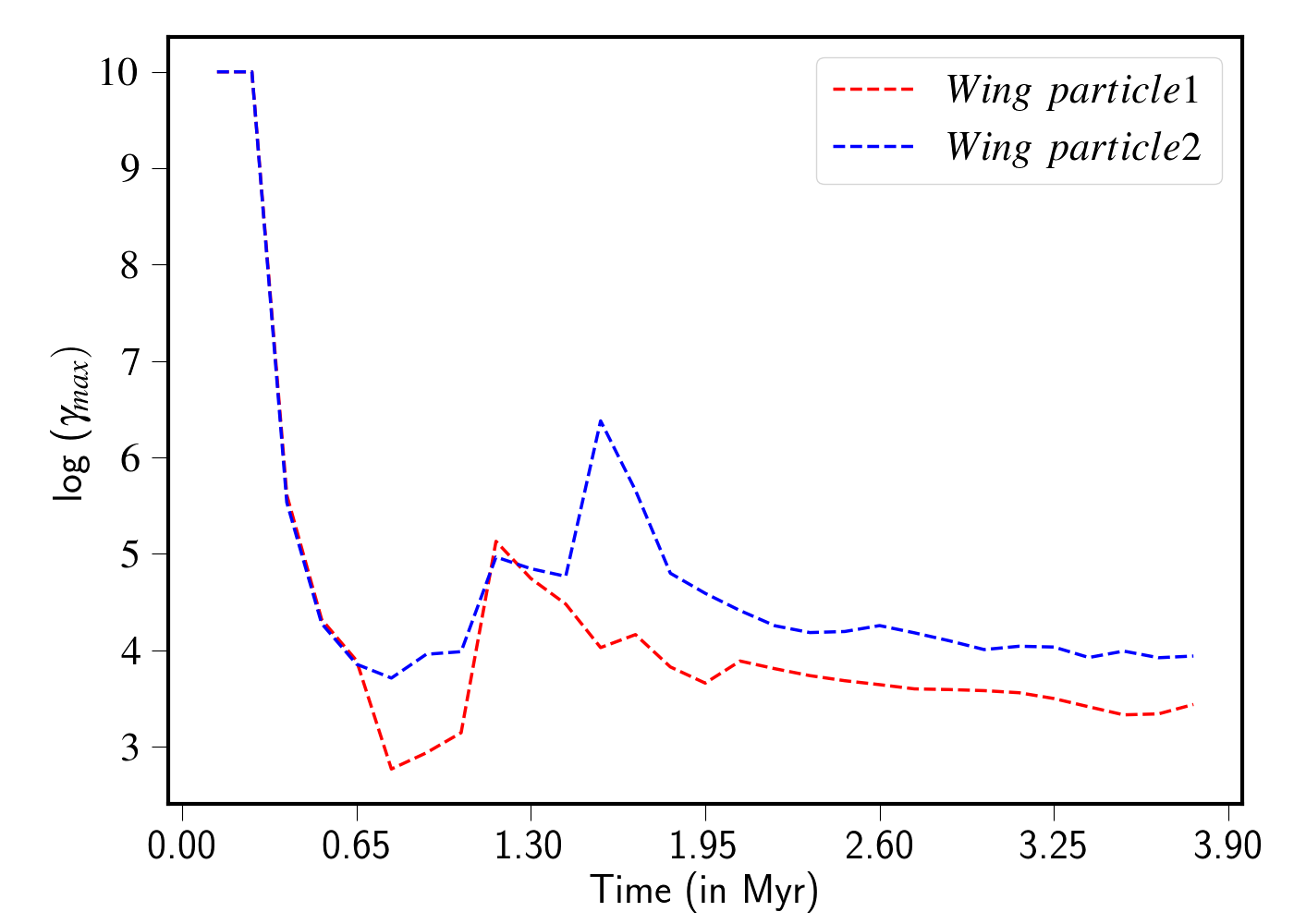}
\caption{Time evolution of maximum Lorentz factor ($\gamma_{\rm max}$) of two wing particles that have undergone multiple shocks in their lifetime ($\geq 4$). The corresponding compression ratios suggest that the shocks are mostly moderate to weak in strength, which is responsible for the secondary peak as shown above. Despite this, the final $\gamma_{\rm max}$ values have reduced significantly from their initial stage.}
\label{Fig:gm_ev}
\end{figure}

We further show explicitly the effect of re-energization via diffusive shocks on particle spectra of two wing particles in Fig. \ref{Fig:gm_ev}. The figure represents the time evolution of $\gamma_{max}$ values for such wing particles which have undergone multiple shocks ($\geq$ 4) in their lifetime. The strength of these shocks are found to be of moderate to weak (i.e. compression ratio $<$ 3). We see an initial steep fall of $\gamma_{max}$ values from $10^{10}$ (initial) to $10^3$ - $10^4$ due to the dominant cooling effects. However, after a time $\sim$ 0.7 Myr, these particles undergo multiple shocks that results in a secondary peak observed in Fig. \ref{Fig:gm_ev}. The cooling processes affect the radio galaxy throughout the time and so, once the particles cross the shock locations, they again start to cool down. As a result of which the $\gamma_{max}$ values again start to reduce after a time of $\sim$ 1.3 Myr and 1.7 Myr for particle 1 (red curve) and 2 (blue curve) respectively. The cooling is mostly governed by radiative process, however, the role of adiabatic effect can not be ignored. The small scale dips found in Fig. \ref{Fig:gm_ev} are mainly due to dominant adiabatic cooling. From this exercise, we elaborately show the significance of diffusive shocks in pausing the drastic cooling of radiating electrons in the wings that keep the structure active despite their ageing with time.

\subsection{On Equi-partition Approximation} \label{On Equi-partition Approximation}
An imprint of particle acceleration due to diffusive shocks is also reflected on the equipartition approximation. While measuring the strength of the magnetic field from direct observation of synchrotron radiation of radio galaxies, it is often assumed that the radiating electrons are in equipartition with the magnetic energy density of the cocoon. This minimum energy condition has also been applied to XRGs while quantifying several physical aspects of it \citep{Hodges-Kluck2010,Cotton2020}. 
The equipartition condition can be written in cgs units as \citep{Hardcastle2002}  
\begin{equation}
U_e = m_e c^2 \ \int_{\gamma_{\rm min}}^{\gamma_{\rm max}} \gamma N(\gamma) d\gamma = \frac{B_{\rm eq}^2}{8\pi}
\label{eq:equip}
\end{equation}
where $U_e$ represents the electron energy density, $N(\gamma)d\gamma$ is the non-thermal electron number density between Lorentz factor $\gamma$ to $\gamma + d\gamma$, $m_e c^2$ is the rest mass energy of electron and $B_{\rm eq}$ is the equipartition magnetic field strength. 
Using Eq~\ref{eq:equip}, we have evaluated the $B_{\rm eq}$ for the lobe and wing particles separately (without considering overlapping region; Section \ref{Spectral Maps and Energy Distribution}). 

\begin{figure}
\centering
\includegraphics[width=\columnwidth]{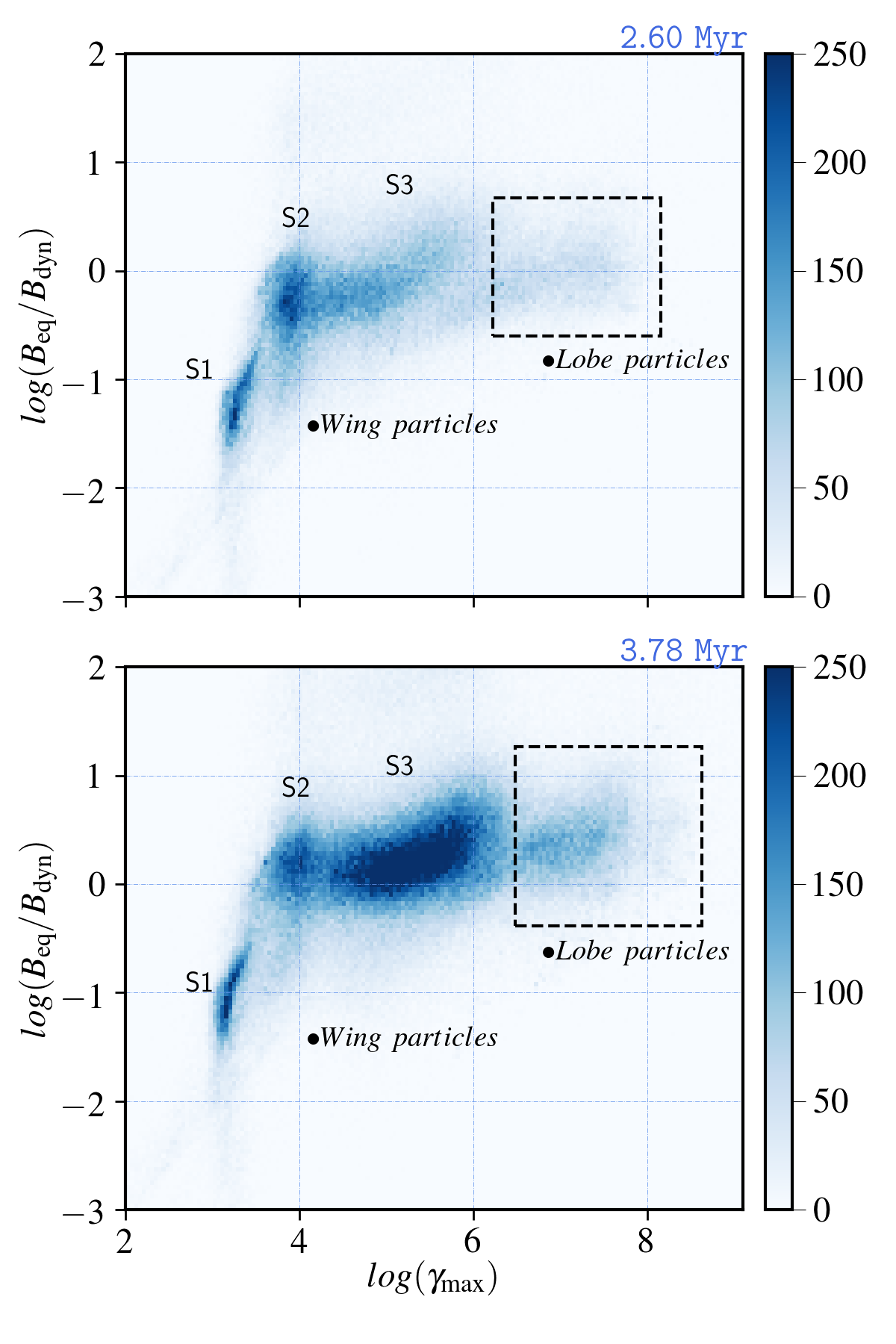}
\caption{2D PDF of $B_{\rm eq}/B_{\rm dyn}$ vs $\gamma_{\rm max}$ of the particles that are in the lobe and in the wing, obtained at a time 2.6 Myr (\textit{top}) and 3.78 Myr (\textit{bottom}). The particles inside the black dashed box are associated with lobes, outside of which, it is the wing particles. The figure shows that most of the particles has acquired a near equipartition stage i.e. $B_{\rm eq} \approx B_{\rm dyn}$, while a few have also deviated from it. Stripe 1, 2, 3 (i.e. S1, S2, S3) marked in the figure are associated with three types of particles that are found in the wings which indicate several micro-physical processes occurring in the wing (see Section \ref{On Equi-partition Approximation} for details). The corresponding colorbar represents the number of particles that are distributed in the bins where we put limit for visualization.}
\label{Fig:eqp}
\end{figure}

In Fig.~\ref{Fig:eqp}, we have represented a 2D PDF showing the distribution of $\frac{B_{\rm eq}}{B_{\rm dyn}}$ vs $\gamma_{\rm max}$ for these particles at time 2.6 Myr (top) and 3.78 Myr (bottom). The particles that belong to the lobes are marked within the black dashed box while the rest belong to the wing. 
The dynamical magnetic field evaluated at the particle's position is denoted by $B_{\rm dyn}$ such that $B_{\rm dyn} = B_{\rm eq}$ represents equipartition. All particles are injected to have an initial sub-equipartition condition (i.e. $B_{\rm eq} = 0.01B_{\rm dyn}$) (see Section~\ref{Emission model}). 
Fig.~\ref{Fig:eqp} evidently shows particles that exist within the wings can be  distinguished into three stripes viz., S1, S2 and S3.

Let us first consider the figure plotted at time 3.78 Myr (Fig. \ref{Fig:eqp}). Particles associated with stripe S1 are the oldest of particles injected into the domain at a very early time and have now radiated away all their energy resulting in $\gamma_{\rm max} \sim 10^{3}$. Majority of the wing particles belong to stripes S2 and S3 and have acquired a near equipartition stage i.e. $B_{\rm eq} \approx B_{\rm dyn}$. 
Their $\gamma_{\rm max}$ values ranges from 10$^{4}$ to 10$^{6}$. The wing particles that belong to stripe S3 have synchrotron cooling time $\leq$ 1.7 Myr for an average B-field strength of 12 $\mu$G obtained for the radio galaxy. This estimated time is an upper limit as particles have also spent some time of their journey in the lobes where B-field strengths are typically higher. We have verified that the majority of particles in stripe S3 have spent more than 1.7 Myr since their injection. In absence of any particle acceleration process, one would have expected these particles to cool and populate the S1 stripe with lower $\gamma_{\rm max}$ and $B_{\rm eq} < B_{\rm dyn}$, instead these particles have near equipartition field strength implying that shock acceleration in the turbulent wings have played a role in altering their non-thermal radiation through re-energization. 
Such re-energization of the non-thermal electrons is mostly due to moderate to weak shocks as seen in our simulations.
The particles associated with  stripe S2 are injected in the domain around a simulation time of 1.4 Myr. They consists of both the cooling and shocked populations. The shocked particles have undergone through multiple shocks in their initial stage of evolution, but now are cooling down without any recent acceleration event (see Fig. \ref{Fig:gm_ev}). 

Further, to illustrate that the above mentioned mechanisms maintain a population of high energetic particles in the wing with time, we now compare the plot shown at time 3.78 Myr with the plot at 2.60 Myr of Fig. \ref{Fig:eqp}. Considering the wing particles, we see their equipartition condition upgrading to higher values with time. In the S1 patch however, the $\gamma_{\rm max}$ values reduce with time (contrary to S2 and S3 patch) indicating that it is only the reduction of dynamical magnetic field value ($B_{\rm dyn}$) in the cocoon (due to the adiabatic expansion) that governs its equipartition condition. These are the oldest injected particles in the domain and are now cooling down under the dominant cooling effects. So, we see particles with limited/no re-acceleration events would exhaust their energy via cooling resulting in $\gamma_{\rm max} \sim 10^3$, and hence emission in $\sim$ MHz range (critical synchrotron frequency).  In this regard, the S2 and S3 patch particles show a systematic shift towards higher $\gamma_{\rm max}$ values with time, as well as the patches get diffuse with greater number of particles, indicating continuous filling of high-energetic particles in the wings with time. Particle re-acceleration via diffusive shocks (as discussed earlier) and particle mixing of different aged populations \citep{Mukherjee2021} are the two leading mechanisms behind the observed characteristics of these patches, which plays a crucial role in maintaining a high energetic particle population in the wing. Since wing particles show a distribution peaking at $\gamma_{\rm max} \sim 10^{4.5}$ (Fig. \ref{Fig:shk}), they would strongly contribute to higher radio frequencies ($\sim$ GHz range) and hence flatten the spectrum as we obtain.

Our analysis regarding the evolution of equipartition condition also applies to the radio galaxies in general. The evidence of it can be found in Fig. \ref{Fig:eqp} which shows that the majority of lobe particles have acquired the equipartition state, although they have been injected into the numerical domain with an initial sub-equipartition condition (Section \ref{Emission model}). Although the number of shocks encountered by the lobe particles are less compared to the wing particles (Fig. \ref{Fig:shk}), the strength of these shocks are high, which is one of the reasons for upgrading the equipartition condition of these particles. At the same time, the cooling also affect the particle spectra as can be seen in Fig. \ref{Fig:gm_ev}, which are capable of reducing the initial $\gamma_{\rm max}$ value drastically from $10^{10}$. Governed by the combined effect of cooling and re-energization, the equipartition condition of the non-thermal particles in lobe will evolve and hence its effect on spectral age will be imparted (discussed further in this section). We note here that the ongoing supersonic turbulence due to the diverted back flow from the lobe is one of the prime mechanisms responsible for re-energizing the wing particles. In the lobes, however, particles can get energized at the sites of recollimation, termination, shear interface of the jet, as well as in the turbulent back-flows,  \citep{borse2021,Mukherjee2021}. 

\begin{figure}
\centering
\includegraphics[width=\columnwidth]{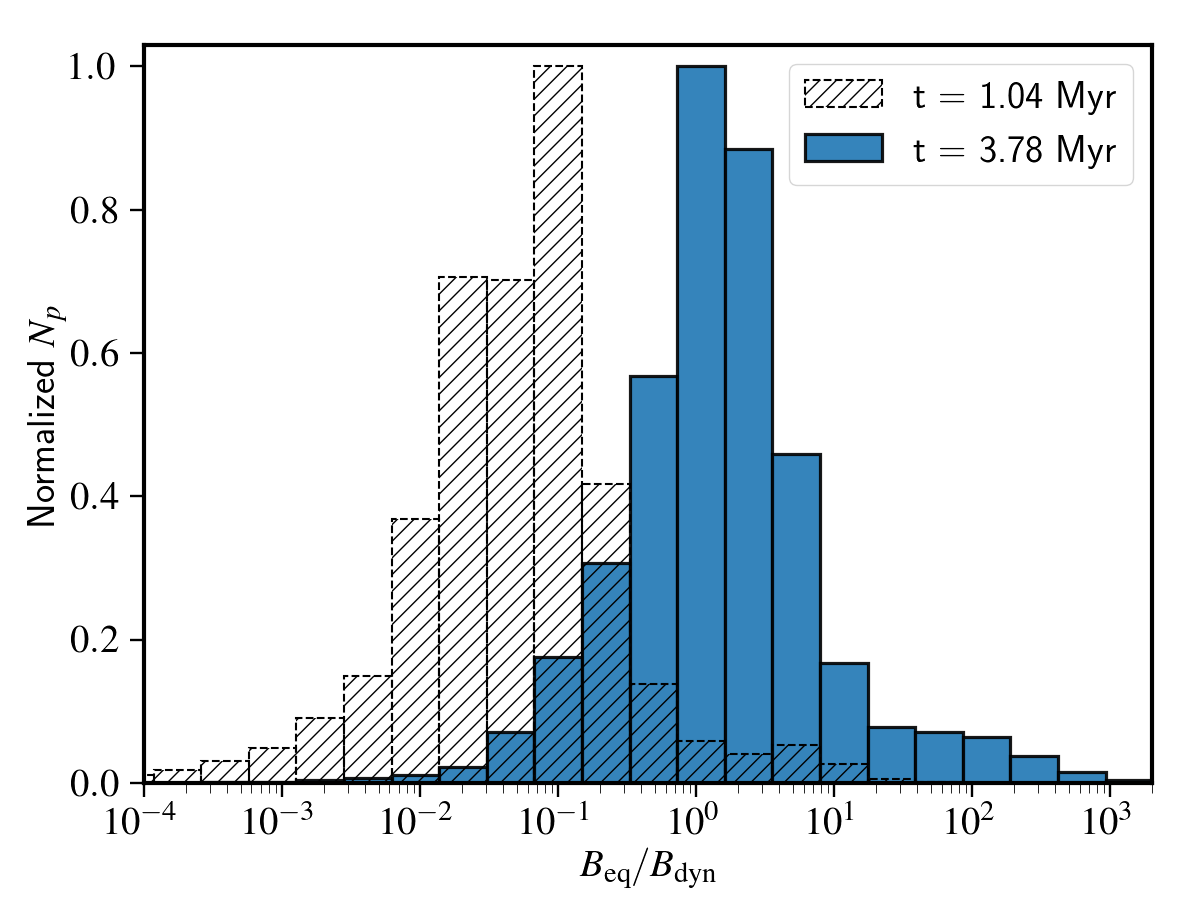}
\caption{Histograms, showing the distribution of $B_{\rm eq}/B_{\rm dyn}$ for all the particles distributed over the domain at a simulation time (t $=$) 1.04 Myr and 3.78 Myr. The vertical axis represents here the normalized number of particles ($N_{\rm p}$). A clear shift in the peak value of the histogram can be observed here as the galaxy grows old. However, at 3.78 Myr, the majority of particles show that the radio structure as a whole has achieved a near equipartition stage ($B_{\rm eq} \approx B_{\rm dyn}$).}
\label{Fig:histogram}
\end{figure}

As indicated earlier, the equipartition condition also varies with time. Here, we have considered the contribution from all the particles injected into the computational domain (i.e. of the entire structure) and estimated $\frac{B_{\rm eq}}{B_{\rm dyn}}$. This magnetic field ratio at simulation times of 1.04 Myr and 3.78 Myr is represented  as histograms in Fig. \ref{Fig:histogram}. 
Following the initial condition discussed in Section~\ref{Emission model} one would expect a peaked distribution as $B_{\rm eq} = 0.01 B_{\rm dyn}$ at time t = 0. With time, we observe that the distrbution is diffused and also the peak is shifted to higher value of magnetic field ratio. At the final time of 3.78\,Myr, we find that the peak value corresponds to near-equipartition stage ($B_{\rm eq} \approx B_{\rm dyn}$) after a systematic shift of one order of magnitude in its value at 1.04\,Myr.
Additionally at 3.78\,Myr, we do have very few particles with a magnetic field ratio $> 100$ as well. This trend indicate that over time the magnetic field ratio may alter particularly if the system is allowed to evolve for a much longer time.

Synchrotron cooling depends strongly on the B-field, a small deviation from equipartition condition of a radio galaxy can lead to a significant difference between spectral and dynamical age \citep{Mahatma2020}. This in turn can affect the estimation of other parameters (for example the Luminosity, jet kinetic power). The study led by \citet{Hodges-Kluck2010} have also shown disagreement between spectral and dynamical age in the context of an X-shaped radio galaxy. Additionally, \citet{Kraft2005} have shown a spatial variation of equipartition condition for XRG 3C 403 by measuring X-ray emission in lobe and wings which can have sigificant influence on age estimation.  
The disparity in the validity of equipartition condition from observations \citep[see also][]{Croston2005} further assert that this approximation is dynamic and has its own limitations in extending  for estimating age. 

For our case, the dynamical age of the radio structure is obtained as the age up to which the simulation has been conducted (t). The corresponding spectral age for an equipartition magnetic field strength $B_{\rm eq}$ is evaluated as
\begin{equation}
    {\rm t}_{\rm spec} = 50.3\frac{B_{\rm eq}^{1/2}}{B_{\rm eq}^2 + B^2_{\rm CMB}}((1+z)\nu_b)^{-1/2}\ {\rm Myr}
\end{equation}
where $B_{\rm eq}$ is expressed in units of nT, $z$ represents the redshift of the source adopted to be 0.05, $B_{\rm CMB} = 0.318(1+z)^2$ is the magnetic field equivalent to cosmic microwave background radiation in nT and $\nu_b$ is the break frequency in GHz \citep{kardashev1962,Miley1980,Pandge2021}. The spectral age can therefore be estimated using equipartition field strength ($B_{\rm eq}$) and break frequency $\nu_b$. 
For time t $=$ 1.04 Myr, we obtain an density weighted average dynamical magnetic field ($B_{\rm dyn}$) of the radio galaxy as 8.4 nT. 
This corresponds to an equipartition field strength ($B_{\rm eq}$) of 0.5 nT. This is estimated from Fig. \ref{Fig:histogram} using the obtained median value of 0.06 of the $\frac{B_{\rm eq}}{B_{\rm dyn}}$ distribution. 
We also find no significant break in the emission spectrum of our galaxy up to the frequency 43 GHz at this early simulation time. Assuming 43 GHz as the break frequency, the spectral age is obtained as 14.2 Myr, giving $\frac{\rm t_{\rm spec}}{\rm t} \sim 13.7$. This estimate is an upper limit as the spectral age would be slightly less than the obtained value for a higher $\nu_b$. 
At time t $=$ 3.84 Myr, we get $\frac{\rm t_{\rm spec}}{\rm t} \sim 2.5$ for an average $B_{\rm dyn}$ field strength of 1.2 nT ($B_{\rm eq} = 1.393 B_{\rm dyn}$, 1.393 is the median value of $\frac{B_{\rm eq}}{B_{\rm dyn}}$ distribution at this time (Fig. \ref{Fig:histogram})) and $\nu_b$ of 5 GHz obtained for our structure. This clearly shows that the discrepancy in the dynamical and spectral age reduces noticeably if the equipartition magnetic field strength approaches the dynamical magnetic field strength of the radio galaxy. Our result is consistent with the findings of \citet{Mahatma2020}. The discrepancy factor of 2.5 (obtained when the galaxy is in near-equipartition state) between the spectral and dynamical age can be attributed to the mixing of different aged particle populations with different evolution histories as suggested by \citet{Mahatma2020}. 
A signature of this blending in our case has been demonstrated in Fig. \ref{Fig:eqp} (see Section \ref{Results 3D: Particle Properties}). The mixing of different particle population is expected due to turbulence particularly at the shear interface which can alter the total particle spectrum when averaged over a particular area \citep{Mukherjee2021}. Such a change in particle spectra can affect the spectral age and results in a factor of a few discrepancy in the spectral/dynamical age discrepancy as has been demonstrated by \citet{Turner2018}. 

One thing to note here is that our simulated radio structure is still in its initial stage of evolution and the near equipartition stage achieved for the radio galaxy is at a dynamical time of 3.78 Myr. By looking at the evolution trend of $\frac{B_{\rm eq}}{B_{\rm dyn}}$, we expect the ratio to peak at a higher value if the structure evolves for a much longer time as also stated earlier. This particularly will raise a spectral age of the radio galaxy which would be less than its dynamical age, typically the case found in observational studies. However the exact evolutionary behaviour of $\frac{B_{\rm eq}}{B_{\rm dyn}}$ in the longer time regime is yet to be explored and would be taken up for further studies.

In summary, our results and analysis indicate that the equipartition condition is rather a dynamic process that not only varies with time but also has spatial dependence within the radio galaxy. Acceleration of particles due to multiple shocks in the turbulent wing structures and also in the lobes have a vital role to play in governing the maximum particle energy, particle spectral index and equi-partition field strength and therefore have implications on the age estimation.


\section{Summary} \label{Conclusions}
We have performed relativistic magneto-hydrodynamic simulations of jet propagation from a tri-axial galaxy to understand the formation of X-shaped radio galaxies based on the Back-flow model. 
We affirm the role of pressure gradient of the ambient and hence the jet ejection direction in forming the X-shaped morphology as suggested by \citet{Capetti2002}. We have considered two principle cases from \citet{Rossi2017} that involves the jet being ejected along the major axis of the galaxy and at an angle $30^{\circ}$ to the major axis of the galaxy in order to investigate their dynamical and emission perspective in detail. To understand the effect of back-flow on the emission signatures, we have modelled the non-thermal emission using a hybrid approach considering the role of the diffusive shock acceleration (DSA) and the radiative and adiabatic cooling on particle spectral evolution. Using such an hybrid approach, one can get a detailed understanding of radiative processes in wings and lobe through the synthetic emission maps which is missing from the previously adopted fixed spectrum approach by \citet{Rossi2017}. Further, the adopted hybrid approach including the effects of diffusive shock acceleration allows us to comprehend the impact of shocks in particle energetics and spectral signatures particularly in the wing region that is typically comprising of older particles. 

The main results from the present work can be summarized as follows:
\begin{itemize}
    \item[$\bullet$] \textbf{Lobe and Wing Formation:} The spatial extent of the wing and the lobe depends strongly on the direction of jet flow. When the jet travels along the major axis of the galaxy, broader wings with prominent X-shape appear with it. Whereas, when it travels at an angle to the major axis of the galaxy, the prominence of the formed structure decreases due to the appearance of higher lobe length for this case. Moreover, we see asymmetry in wing formation i.e. the bending of back-flowing materials before reaching centre for the angled jet case, whose morphological relevance can be found in several observed XRGs. Owing to a value of $\beta_{\rm pl} \gg 1$, the wings expand as thermal pressure dominated flow. 
    
    The lobe length follows the analytical estimates of a 1D jet flow up to 0.9\,Myr and later deviates and lags behind the analytical value due to lateral expansion, consistent with the results from \citet{Rossi2017}. 
    
    \item[$\bullet$] \textbf{Spectral Index and Viewing Angle:}
    The synthetic intensity maps in radio bands obtained with different viewing angles are qualitatively similar in terms of morphology observed in X-shaped radio galaxies. Owing to variation in viewing angle, one also finds synthetic maps with wider wings as compared to the active lobes. 

    The synthetic spectra up to the optical B band (SED) show a typical peaked structure with the spectral peak around 1.4\,GHz as expected due to syncrotron emission. 
    The associate spectral index ($\alpha$) map of the galaxy infers that the wings generally show steeper spectral index distribution owing to the concentration of older particles (sufficiently cooled). However, there exist significant patches of flatter $\alpha$-values that are generated due to re-energized particles at shocks. Moreover, this distribution depends on the choices of frequencies, where for wider choices of it the obtained values get more and more steeper in the wings.
    Based on this and the viewing effect, we have quantified $\Delta\alpha_{\nu1}^{\nu2}$ $=$ $\alpha^{\rm lobe}_{\rm av}$ - $\alpha^{\rm wing}_{\rm av}$ and have found that the Back-flow model is capable of producing values that span both in the $+$ve and -ve side. However, we show that this is true only for frequencies that lie in the rising phase of the SED (up to the peak), whereas, for higher frequencies, the obtained $\Delta\alpha_{\nu1}^{\nu2}$ values are prone to have a $+$ve distribution due to the cooling of particles. Additionally, the projected size (viewing angle) of the structure also influences the trend of $\Delta\alpha_{\nu1}^{\nu2}$ values.
    This degeneracy in the estimate of $\Delta\alpha_{\nu1}^{\nu2}$ value indicates that such a measure is not sufficient to identify and uniquely determine the formation mechanism of XRGs.  
    
    \item[$\bullet$] \textbf{Particle Evolution and Equipartition:}
    The non-thermal electron population, initially distributed according to a power law, undergoes cooling due to the radiative (major influence) and adiabatic losses right after their injection into the domain. In this regard, the diffusive shock acceleration re-energizes these particles and counters the effect of cooling in particle spectra. We show that this has a significant role to play in wing emission where the shocks pause the drastic cooling of particles that eventually keep the wing structure active during the evolution time. We find the accumulation of three types of particles in the wing that together governs its evolution. These are - a) the old cooling particles, b) freshly shocked particles and c) particles that have undergone shocks at their early stage but now are cooling.
    
    Using the hybrid framework, we have also compared equipartition magnetic field $B_{\rm eq}$ with dynamical magnetic fields $B_{\rm dyn}$. During the course of evolution, majority of particles show a systematic shift from an initial sub-equipartition population $B_{\rm eq} < B_{\rm dyn}$ to near equipartition due to consistent evolution of particle distribution in presence of radiative losses from local magnetic fields and diffusive shock acceleration. 
    This analysis indicates that equipartition between the radiating and magnetic energy is rather a dynamic process that evolves with time. Such an evolution has a significant impact on the estimate of age of the radio galaxies and observed discrepancy between spectral and dynamical age \citep{Mahatma2020}. We also have elaborated the effect of different aged particle population mixing in the dynamical and spectral age discrepancy, which one must take into consideration while estimating spectral age from radio observations.
\end{itemize}

The present study showcases the role of back-flow in forming X-shaped radio galaxies along with its emission signature from hybrid modelling of non-thermal particles. The focus of this work has only been on the initial formation phase of XRGs. Our model demonstrates several characteristic features that are the possible origin of the XRG structures observed at much larger scales. We have seen that the adopted back-flow model with a combined effect of viewing angle and multi-band spectral maps can explain the observed features of XRGs. However, one would need to affirm the universality of this model in formation of XRGs using large scale and long term simulations \citep{Hodges-Kluck2011}. Additionally, it would be interesting to understand the interplay of turbulence and shock acceleration \citep{Kundu2021} in wings and its impact on radiative signature. 
These studies along with adopting competing formation models would be discussed in forthcoming papers.

\begin{acknowledgements}
The authors thank Alessandro Capetti for providing valuable suggestions and comments. GG is supported by Prime Minister's Research Fellowship and thank the financial support provided. BV is leading a Max Planck Partner Group at Indian Institute of Technology Indore would like to thank The Council of Scientific \& Industrial Research (CSIR) [03(1435)/18/EMR-II]. Numerical simulations presented in this work are carried out using facilities at Indian Institute of Technology Indore and MPG Super-computing Cobra cluster \url{https://www.mpcdf.mpg.de/services/supercomputing/cobra}.
\end{acknowledgements}

%
%


\bibliographystyle{mnras} 
\bibliography{sample} 
\end{document}